\documentclass[fleqn,usenatbib]{mnras}

\usepackage[T1]{fontenc}
\usepackage{ae,aecompl}

\usepackage{graphicx}

\usepackage{amsmath}

\numberwithin{equation}{section}
\renewcommand{\thesection}{\arabic{section}}

\makeatletter

\usepackage{bm}
\usepackage{graphicx}	
\usepackage{amsmath}	
\usepackage{amssymb}	
\usepackage{adjustbox}
\usepackage{comment}
\usepackage{siunitx}
\usepackage{float}
\usepackage{xspace}
\usepackage{etoolbox}
\usepackage{multirow}
\usepackage{tcolorbox}

\usepackage{newtxtext,newtxmath}

\title[Searching for Intelligent Life in Gravitational Wave Signals Part I]{Searching for Intelligent Life in Gravitational Wave Signals Part I: \\ Present Capabilities and Future Horizons}

\author[L.Sellers et al.]{Luke Sellers$^{1,2}$\thanks{Corresponding author: luke@appliedphysics.org},
Alexey Bobrick$^{1,3,4}$,
Gianni Martire$^{1}$,
\newauthor
Michael Andrews$^{5}$,
Manfred Paulini$^{5}$
\\
$^{1}$Advanced Propulsion Laboratory at Applied Physics, 477 Madison Ave, New York, 10022, United States\\
$^{2}$University of California Los Angeles, Department of Physics, Los Angeles, CA, 90095, USA\\
$^{3}$Technion - Israel Institute of Technology, Physics Department, Haifa 32000, Israel\\
$^{4}$Lund University, Department of Astronomy and Theoretical Physics, Box 43, SE 221-00 Lund, Sweden\\
$^{5}$Carnegie Mellon University, Department of Physics, Pittsburgh, Pennsylvania 15213, USA\\
}

\date{Accepted XXX. Received YYY; in original form ZZZ}

\pubyear{2022}

\begin{document}
\label{firstpage}
\pagerange{\pageref{firstpage}--\pageref{lastpage}}
\maketitle


\definecolor{awesome}{rgb}{1.0, 0.13, 0.32}


\begin{abstract}
We show that the Laser Interferometer Gravitational Wave Observatory (LIGO) is a powerful instrument in the Search for Extraterrestrial Intelligence (SETI). LIGO's ability to detect gravitational waves (GWs) from astrophysical sources, such as binary black hole mergers, also provides the potential to detect extraterrestrial mega-technology, such as \textbf{R}apid \textbf{A}nd/or \textbf{M}assive \textbf{A}ccelerating spacecraft (\textsc{RAMAcraft}). We show that LIGO is sensitive to \textsc{RAMAcraft} of 1 Jupiter mass accelerating to a fraction of the speed of light (e.g. 30\%) from $10 - 100\,$kpc or a Moon mass from $1-10\,$pc. While existing SETI searches can probe on the order of ten-thousand stars for human-scale technology (e.g. radio waves), LIGO can probe all 10$^{11}$ stars in the Milky Way for \textsc{RAMAcraft}. Moreover, thanks to the $f^{-1}$ scaling of \textsc{RAMAcraft} signals, our sensitivity to these objects will increase as low-frequency detectors are developed and improved, allowing for the detection of smaller masses further from Earth. In particular, we find that DECIGO and the Big Bang Observer (BBO) will be about 100 times more sensitive than LIGO, increasing the search volume by 10$^{6}$, while LISA and Pulsar Timing Arrays (PTAs) may improve sensitivities to objects with long acceleration periods. 
In this paper, we calculate the waveforms for linearly-accelerating \textsc{RAMAcraft} in a form suitable for LIGO, Virgo, and KAGRA searches and provide the range for a variety of masses and accelerations. We expect that the current and upcoming GW detectors will soon become an excellent complement to the existing SETI efforts.
\end{abstract}





\section{Introduction}
\vspace{-0.1cm}


Gravitational wave (GW) detectors such as LIGO and Virgo \citep{Abbott_2020_noise} have contributed three entirely new signals to the astronomical catalog, namely GWs from binary black hole (BBH) mergers \citep{PhysRevLett.116.061102}, neutron star-black hole (NSBH) mergers \citep{BHBS_inspiral_detection}, and binary neutron star (BNS) mergers \citep{binary_inspiral_detection}. Each of these signals consists of a mutual orbit of stellar-mass (1 -- 100$\,{\rm M}_\odot$), compact objects that LIGO and Virgo have detected as far out as several thousand $\,{\rm Mpc}$. The accumulation of these detections has led to an array of significant findings, including stringent tests of general relativity \citep{strong_field_tests}, a novel method for determining the cosmological constant \citep{hubble}, and evidence of various neutron star processes, such as their connection to gamma-ray bursts \citep{binary_inspiral_detection}, their equation of state \citep{NS_eq_of_state}, and the production of heavy r-process elements, such as gold and uranium \citep{NS_metals}. Furthermore,  the detection of new signals in the future will shed light on topics including black hole (BH) formation, the quantum behavior of BHs, structure formation in the early universe, dark matter, and dark energy
\citep{nature_GW_uses, fundphys_lisa, uofv_thesis, axion_gwaves, Barack_2019, dark_energy_constraints}.
\par
The detection of these new signals will be dramatically enhanced by the sensitivity improvements provided by future GW detectors. These improvements will come in two types, namely the improvement of overall detector sensitivities and the broadening of detector sensitivity bands. In particular, the planned Einstein Telescope \citep{ET} and Cosmic Explorer \citep{CE} will improve the overall sensitivity, leading to the discovery of thousands of BBH mergers from across the observable universe each day. At the same time, the upcoming detectors LISA \citep{LISA_sens}, TianQin \citep{TianQin}, DECIGO \citep{DECIGO}, the Big Bang Observer (BBO) \citep{BBO}, and improvements to Pulsar Timing Arrays (PTAs) \citep{PTAs, realistic_curves_PTA}, such as the Square Kilometer Array (SKA) \citep{SKA}, will collectively widen the sensitivity band of the detector network. This will lead to the detection of entirely new sources, such as supermassive BH mergers, extreme-mass-ratio inspirals, close white dwarf and subdwarf binaries, amongst others. Collectively, these improvements will vastly extend the number of detected signals and lead to the discovery of multiple new types of astrophysical sources.
\par 
Since any system involving the bulk acceleration of mass produces GWs, new signal candidates include not only astrophysical and cosmological events, but also technological signals (technosignatures), such as those generated by rapid and/or massive accelerating spacecraft (\textsc{RAMAcraft}). We can already perform targeted searches for any such object simply by calculating its GW signal and plugging the result into existing detection pipelines. In this paper, we take a first look at the capabilities of LIGO and future GW detectors, such as LISA, to detect these objects. Specifically, we compute the range at which a linearly-moving \textsc{RAMAcraft} with constant acceleration will be detectable by either LIGO or LISA. In the case of LIGO, we set novel constraints on the occurrence of such trajectories while LIGO was online. 
\par
Though our focus is on artificial objects, it may be the case that the signals that we examine here are mimicked by some naturally occurring phenomena.
In Section~\ref{sec:Discussion}, we discuss a potential candidate, namely highly-eccentric BBH mergers. However, BBH mergers with extreme eccentricities are expected to be quite rare and are not particularly likely to occur in their own right. 
\par
In the artificial case, a detected GW signal may be the result of an advanced transportation mechanism utilized by an intelligent species. Thus far, the search for such objects has primarily been restricted to the examination of electromagnetic (EM) signals through the efforts of the Search for Extraterrestrial Intelligence (SETI) \citep{SETI_2020, breakthrough_listen, BL_TESS} and like-minded research groups \citep{loeb_subrel_2020, Oumuamua_2018}. For example,  it has been suggested that thermal emission due to the surface heating of a 100 m-scale interstellar spacecraft could be detected by the James Webb Space Telescope (JWST) from up to about $100\,{\rm AU}$ \citep{loeb_subrel_2020}. Similarly, multi-color telescopes can also detect 100 m-scale interstellar objects (ISOs) within our solar system, which was famously demonstrated in 2017 with the detection of `Oumuamua' \citep[][although the latter is not necessarily of an artificial origin]{Oumuamua_2018}.
\par 
Adding a GW detection pipeline for \textsc{RAMAcraft} can provide several important benefits in the search for extraterrestrial life. First, not all systems emit significant EM waves. Second, GW detectors have a vast field of view, while EM images comprise a small fraction of the sky. Third, due to the weakness of gravitational interactions, all GW signals of all frequencies travel through space virtually unimpeded. By contrast, substantial EM radiation absorption occurs, for example, in the interstellar medium for low-frequency radio, optical, and ultraviolet wavelengths and ultraviolet, microwave, and X-ray bands in the Earth's atmosphere. Consequently, detecting EM signals for certain frequencies may be challenging or require space-based telescopes. The fourth advantage is that the GW quantity that is observed, namely the \textit{strain}, falls off like $R^{-1}$ where $R$ is the distance to the source, whereas the observed EM quantity, namely the \textit{flux}, falls off like $R^{-2}$. 
This difference in scaling can be significant over large distances, enabling the upcoming detectors to probe the entire observable universe for BBH mergers, for example. Finally, GW detectors monitor natural and SETI sources simultaneously. By contrast, EM searches typically require pointing telescopes at one target at a time, thereby competing with the observation time needed for natural sources and other SETI targets. Despite these difficulties, SETI and other like-minded research groups have extended EM searches to the order of tens of thousands of stars. For example, one of SETI's most successful programs, the Breakthrough Listen Initiative \citep{breakthrough_listen}, set constraints on optical signatures from advanced civilizations for 1,327 stars through a recent search \citep{Price_2020}.  Furthermore, a SETI press release announced in 2016 a search for 20,000 stars. Although these feats are remarkable, extending these searches by orders of magnitude will be challenging due to the aforementioned difficulties involved with EM searches. 
\par 
For these reasons, we propose searching for \textsc{RAMAcraft} and other technosignatures by studying GW signals. 
Similar to the targets of astrophysical GW searches, the
prime targets for GW SETI searches at the present moment must be objects from so-called mega-technology, which includes objects of planetary-scale mass and/or with changes in velocity comparable to the speed of light (e.g. $10\rm \%$). While these parameter scales go beyond anything familiar to us on Earth, systems of interest such as physical warp drives \citep{jackson2020novel,BobrickMartire21}, Dyson spheres \citep{dyson_attenuation, dyson_gaia}, and others necessarily involve such masses. In a similar spirit, the study by \citet{Gourgoulhon_2019, nature_2019} showed that an orbiting body of Jupiter-mass scale intended to extract energy from Sagittarius A (Sgr A)  would be detectable by LISA. As we show in Section~\ref{sec:Results}, our detectable parameter space for \textsc{RAMAcraft} is comparable to these results. 
Furthermore, as we show in Section~\ref{sec:Discussion}, what may bridge the gap between these large mass scales and something more familiar for the case of \textsc{RAMAcraft} is that the GW signal spectrum scales like $f^{-1}$ and does so even past the inverse acceleration period due to the gravitational memory effect. Therefore, detectors that are more sensitive to lower frequencies will be more sensitive to these objects.
\par 
This paper is structured as follows: In Section~\ref{ref:Sec2LinAcc}, we calculate the time-domain signal produced by a RAMAcraft operating as a Newtonian rocket, then calculate the frequency spectrum, focusing on the case of constant acceleration. In Section~\ref{sec:Results}, we compute the distance from Earth at which such an object would be detectable by both LIGO and LISA for a given mass and change in velocity. In Section~\ref{sec:Discussion}, we discuss how detection feasibility can be improved by future detector improvements and how this approach can be generalized for future studies. Finally, we offer concluding thoughts in Section~\ref{sec:Concl}. 
\section{Gravitational Wave Signal from Linearly-Accelerating Sources}
\label{ref:Sec2LinAcc}
In this section, we calculate the time-domain GW signal emitted by an object with a general linear trajectory. We then apply conservation of momentum ad hoc and find that our result agrees with the exact calculation using a Newtonian rocket, ignoring the effects of exhaust. We show how to account for exhaust and move on to focus on the frequency spectrum of objects undergoing constant acceleration. 
\subsection{Time-Domain Signal}
In general relativity, if the gravitational field is weak, then the metric $g_{\mu\nu}$ that describes the curvature of spacetime can be decomposed into the metric that describes flat space $\eta_{\mu\nu}$ and that which describes a perturbation of flat space $h_{\mu\nu}$ as \citep{carroll}
\begin{equation}
    g_{\mu\nu} = \eta_{\mu\nu} + h_{\mu\nu} 
\end{equation}
where the notion that the field is weak is communicated by the condition that $h_{\mu\nu} \ll 1$ in natural units. The metric perturbation constituting a GW in the Trace-Reversed Lorenz gauge for far away, slow-moving matter is then given by the quadrupole formula:
\begin{equation}
\label{eqn:quad_form}
\overline{h}_{ij}
=
\frac{2G}{Rc^{4}}
\frac{d^{2}I_{ij}}{dt^{2}}
\end{equation}
where $G$ is Newton's constant, $c$ is the speed of light, $R$ is the distance from the source to the detector, and $I_{ij}$ is the quadrupole of the mass distribution:
\begin{equation}
\label{eqn:quad_tensor}
    I_{ij}(t)
    =
    \int{x_{i}x_{j}T^{00}(t,x,y,z)d^{3}x}
\end{equation}
where $T^{00}$ is the energy density of the system and $x_{i}$ is a Cartesian coordinate relative to some origin. We will see later that the choice of origin is inconsequential. 
\par 
It can then be shown that, in the Transverse-Traceless (TT) gauge, the metric perturbation $h_{\mu\nu}$ satisfies the wave equation:
\begin{equation}
    \Box h^{TT}_{\mu\nu} = 0
\end{equation}
where $\Box = -\partial_{t}^{2} + \partial_{i}\partial^{i}$ is the flat space wave operator. In this gauge, $h^{TT}_{\mu\nu}$ has two degrees of freedom, or polarizations:
\begin{equation}
\label{eqn:tt_matrix}
    h^{TT}_{\mu\nu}
    \sim
    \begin{pmatrix}
    0&0&0&0\\
    0&h_{+}&h_{\times}&0\\
    0&h_{\times}&-h_{+}&0\\
    0&0&0&0
    \end{pmatrix}
\end{equation}
where $h^{TT}_{\mu\nu}$ takes the form (\ref{eqn:tt_matrix}) for a wave
\vspace{0.1cm}
propagating in the $\hat{z}$-direction. Then using the geodesic deviation equation, it can be shown that the effect of the wave passing through a group of test masses is to cause them to expand and contract in a `plus' or `cross' pattern, depending on the polarization $h_{+}$ or $h_{\times}$. The signal that is then perceived by the detector is
\begin{equation}
\label{eqn:resp_funcs}
    h_{\rm det}(t) = F_{+}h_{+} + F_{\times} h_{\times}
\end{equation}
where $F_{+}$ and $F_{\times}$ are the detector response functions that depend on the orientation between the source and the detector \citep{antenna_response1}. 
\par
\subsubsection{Signal Calculation}\hfill \\
\par
For simplicity, we define the trajectory $z(t)$ of the object to be in the purely $z$-direction. Then, modeling the object as a point mass, the energy density for an at most mildly relativistic object is given by
\begin{equation}
\label{eqn:en_density}
    T^{00}(t,x,y,z)
    =
    M(t) \cdot 
    \delta(x)\delta(y)
    \delta\left(
    z - z(t)
    \right)
\end{equation}
where $M$ is the instantaneous mass of the object. Note that the choice of origin in (\ref{eqn:en_density}) corresponds to taking the origin in (\ref{eqn:quad_tensor}) as the initial position of the object. The effect of shifting this origin to some inertial frame, such as the center of mass frame, is to change the quadrupole by terms that are at most quadratic in $t$, thereby contributing a non-detectable overall constant to the strain (\ref{eqn:quad_form}) \citep[for details, see Section~10.5 of][]{weinberg}. 
\par 
We then find that all of the components of $I_{ij}$ are zero except
\begin{equation}
    I_{zz}
    =
    M(t)\cdot
    z^{2}(t)
\end{equation}
and that the strain for a general linear trajectory is given by
\begin{equation}
\label{eqn:gen_hzz}
   \overline{h}_{zz}
   =
   \frac{2G}{Rc^{4}}
   \left\{
   2M\dot{z}^{2}
   +
   4\dot{M}z\dot{z}
   +
   2Mz\ddot{z}
   +
   \ddot{M}z^{2}
   \right\}
\end{equation}
where a dot denotes differentiation with respect to the detector frame time $t$. 
\par
At this point, the astute reader will likely protest that such an object ignores the cause of the acceleration and therefore does not constitute a closed system. However, we can just apply the requisite conservation laws ad hoc by modeling our system as a discrete rocket and invoking Newtonian arguments. 
\par 
Suppose that our object achieves its acceleration by ejecting mass at successive points in time. Then by conservation of mass, momentum, and Newton's Third Law at each ejection, it is necessarily the case that
\begin{enumerate}
    \item $\dot{M}_{R} = -\dot{M}_{E} 
    \to
    \ddot{M}_{R} = -\ddot{M}_{E}$
    \item $M_{R} \ddot{z}_{R}= - M_{E} \ddot{z}_{E}$
    \item
    $\dot{M}_{R}\dot{z}_{R} = -\dot{M}_{E}\dot{z}_{E}$
\end{enumerate}
where quantities with subscript $R$ and $E$ pertain to the rocket and exhaust, respectively. Note that the third condition follows from the product rule combined with conservation of momentum and Newton's Third Law.
We are then left with
\begin{equation}
\label{eqn:h_overline_zz}
   \overline{h}_{zz}
   =
   \frac{4GM(t)}{Rc^{4}}
   v(t)^{2}
\end{equation}
where $v=\dot{z}$. For an object with initial velocity $v_{0}$ that undergoes a constant acceleration $A$ at $t=0$ for a period $T$, we have
\begin{equation}
    v(t)
    =
    \begin{cases}
    v_{0} & t<0\\[0.15cm]
    At + v_{0} & 0 < t < T\\[0.15cm]
    AT + v_{0} & T < t
    \end{cases}
\end{equation}
and the strain is given by
\begin{equation}
\label{eqn:h_overline}
    \overline{h}_{zz}
    =
    \begin{cases}
    \frac{4GM}{Rc^{4}}
   v_{0}^{2} & t<0\\[0.15cm]
    \frac{4GM}{Rc^{4}}\left(At + v_{0}\right)^{2} & 0 < t < T\\[0.15cm]
    \frac{4GM}{Rc^{4}}\left(AT + v_{0}\right)^{2} & T < t
    \end{cases}
\end{equation}
Note that the constant velocity contributions do not emit power since the power from GWs is given by $P_{GW} \propto d^{3}I_{ij} / dt^{3}$. Even so, the difference between the two constant velocity contributions should be observable through the GW memory effect \citep{gw_memory}.
\par 
For this study,  we consider objects far enough away and accelerating over sufficiently short distances such that $R$ is effectively constant. This condition is met when
\begin{equation}
\label{eqn:const_R}
    R \gg \frac{1}{2}AT^{2}
\end{equation}
As we can see from the strain (\ref{eqn:h_overline_zz}), objects that accelerate for longer durations are more easily detected, so accounting for the object's motion is well-motivated for future studies.
\par 
We now might like to view the metric perturbation in the TT gauge:
\begin{equation}
\label{eqn:hTT_perp_x}
    h^{TT}_{ij}
    \sim
    \frac{1}{2}
    \overline{h}_{zz}
    \begin{pmatrix}
    0&0&0\\
    0&-1&0\\
    0&0&1
    \end{pmatrix}
\end{equation}
choosing the transverse direction of $n^{i} = \hat{x}$. The form of (\ref{eqn:hTT_perp_x}) means that waves that are emitted in the plane perpendicular to the direction of motion of the object will be purely `+' polarized, in agreement with \citet{rule_of_thumb}.  For a general direction $n^{i}$, we find that
\begin{equation} 
\label{eqn:gen_direction}
h^{TT}_{ij}  = \frac{1}{2}\overline{h}_{zz} \times
\end{equation}
\[
\times \begin{pmatrix}
(n_{x}^{2}+1)n_{z}^{2}+n_{x}^{2}-1 & n_{x}n_{y}n_{z}^{2}+n_{x}n_{y}& n_{x}n_{z}^{3}-n_{x}n_{z} \cr
n_{x}n_{y}n_{z}^{2}+n_{x}n_{y} & (n_{y}^{2}+1)n_{z}^{2}+n_{y}^{2}-1   & n_{y}n_{z}^{3}-n_{y}n_{z} \cr
n_{x}n_{z}^{3}-n_{x}n_{z}& n_{y}n_{z}^{3}-n_{y}n_{z} & (1-n_{z}^{2})^{2}
\end{pmatrix}
\]
Regardless of the normal direction, we see that the polarization components $h_{+}$ and $h_{\times}$ will be on the order of
\begin{equation}
    h_{+,\times} 
    \lesssim
    \frac{1}{2}\overline{h}_{zz}
\end{equation}
with saturation occurring for `+' polarisations of waves emitted in the plane perpendicular to the motion of the object. 
\par 
Since we are striving to attain an order of magnitude estimate of detection ranges for these signals, we will use the form (\ref{eqn:hTT_perp_x}) for the strain and set $F_{+} = 1$. Then the strain measured by the detector is
\begin{equation}
\label{eqn:time_strain}
    h(t) = \frac{1}{2} \overline{h}_{zz}(t)
\end{equation}
For future studies in which the orientation of the source with respect to the detector needs to be examined, using (\ref{eqn:resp_funcs}) and (\ref{eqn:gen_direction}) will be necessary.
\par 
Here we comment on how these results change when the propulsion mechanism is taken into account. We do this by calculating the strain produced by a Newtonian rocket in Appendix \ref{app:app_a}. The cancellations that were justified using Newtonian arguments then occur explicitly, producing (\ref{eqn:h_overline_zz}) and an additional term derived from the exhaust:
\begin{equation}
\label{eqn:h_exhaust}
    \overline{h}_{zz}^{(E)}
    =
    \frac{4G}{Rc^{4}}
    \cdot
    \int_{0}^{t}
    -
    \dot{M}(t')
    u^{2}(t')
    dt'
\end{equation}
where $\dot{M} < 0$ and $u$ is the velocity with which each ejected mass travels, which may change between ejections. In simple terms, the strain contribution from the exhaust is given by its total kinetic energy, just like the rocket contribution. 
\par
\begin{figure}
    \centering
    \includegraphics[width = 1.0\linewidth]{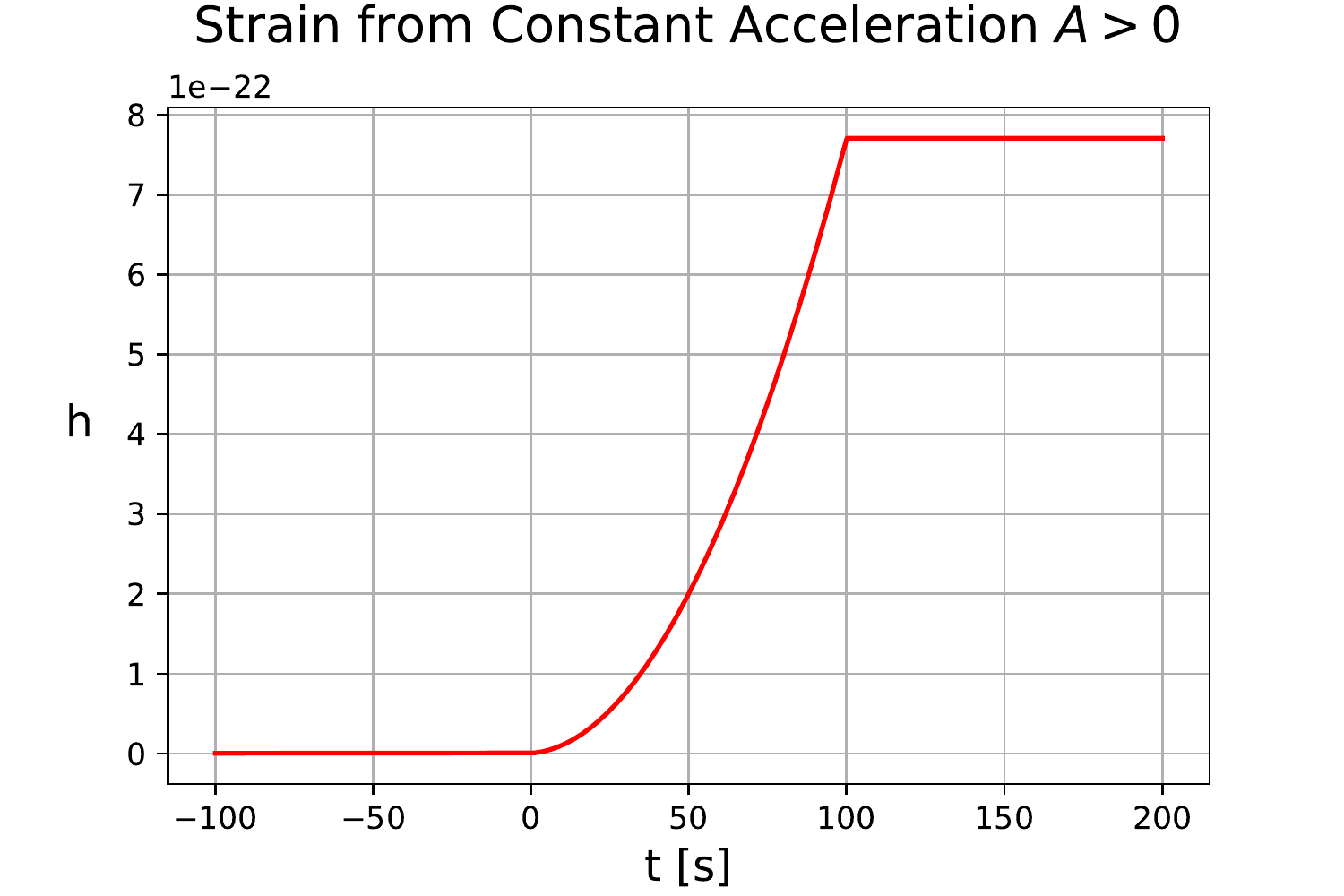}
    \caption{Plot of the GW signal (\ref{eqn:time_strain}) generated by a $10^{-3}\,{\rm M}_{\odot}$ object accelerating constantly from $10^{-2}\,c \to 0.5\,c$ in 100\,s from $10^{21}$\,m ($\approx 30\,{\rm kpc}$) away. The data sample length exceeds the acceleration period, so the constant velocity contributions from (\ref{eqn:h_overline}) are visible. Note that for a more realistic object, the acceleration will not begin or cease instantly, so the beginning and end of acceleration will be rounded off. However, the form (\ref{eqn:h_overline}) suffices for an estimate of detection ranges.}
    \label{fig:1}
\end{figure}

 
The GW signal (\ref{eqn:time_strain}) can be seen in Figure \ref{fig:1} for the case of positive acceleration. We also take the mass $M$ to be a constant and neglect the exhaust contribution, which we do for the remainder of the paper as well. Assuming that the mass is constant is reasonable if the exhaust velocity is sufficiently larger than the change in velocity of the rocket. The sample is then collected between times $t_{1}$ and $t_{2}$. The constant velocity portions then contribute between $[t_{1}, 0]$ and $[T, t_{2}]$. Interestingly, the spectrum generated by positive acceleration is identical to that of negative acceleration if we swap $v_{0} \iff v_{f}$ and $|t_{1}| \iff |t_{2} - T|$. In this manner, we can search for two signals (accelerating and decelerating) at once by conducting our search in the frequency domain. The two cases can still be differentiated by their spectrograms.
\subsection{Frequency-Domain Signal}
For calculating the spectrum of discrete data, we use the Fast Fourier Transform (FFT) convention used in \citet{find_chirp}:
\begin{equation}
\label{eqn:fft_convention}
    h(f_{k})
    =
    \Delta t
    \sum_{n=1}^{N}
    h(t_{n})
    e^{-2\pi i k t_{n} / \tau}
\end{equation}
where $\tau = t_{1} - t_{2}$ is the length of the time-domain data sample. This convention matches the continuous Fourier Transform (CFT) of the signal (\ref{eqn:time_strain}) with $h(t< t_{1})$ and $h(t > t_{2})$ set to zero. Then the discrete spectrum is given by
\begin{equation}
\begin{split}
\label{eqn:fourier_integral}
    \hat{h}(f)
    &=
    \int_{-\infty}^{\infty}{h(t)e^{-2\pi i f t}dt}\\
    &=
    \int_{t_{1}}^{t_{2}}{h(t)e^{-2\pi i f t}dt}
\end{split}
\end{equation}
There are then four cases that are important to consider when calculating (\ref{eqn:fourier_integral}), namely
\begin{enumerate}
    \item 
    \label{case1}The entire acceleration period is captured by the data sample: $t_{1} < 0$ and $T < t_{2}$
    \item 
    \label{case2}
    The initial velocity and part of the acceleration period are captured by the data sample: $t_{1} < 0$ and $0 < t_{2} < T$
    \item 
    \label{case3}
    The final velocity and part of the acceleration period are captured by the data sample: $0< t_{1} < T$ and $T < t_{2}$
    \item 
    \label{case4}
    The data sample is contained entirely within the acceleration period: $0 < t_{1} < T$ and $t_{1} < t_{2} < T$
\end{enumerate}
We then calculate (\ref{eqn:fourier_integral}) for Case \ref{case1}, after which the results are simple to generalize to the other three cases. For this case, the spectrum (\ref{eqn:fourier_integral}) becomes
\begin{equation}
    h(f)
    =
    \frac{2GM}{Rc^{4}} 
    \left(
    I_{1} + I_{2} + I_{3}
    \right)
\end{equation}
where
\begin{equation}
    \begin{split}
        &I_{1} = \int_{t_{1}}^{0}
        v_{0}^{2}e^{-2\pi i f t} dt\\
        & I_{2} = \int_{0}^{T}
        \left(At + v_{0}\right)^{2}
        e^{-2\pi i f t} dt\\
        &I_{3} = \int_{T}^{t_{2}}
        \left(AT + v_{0}\right)^{2}
        e^{-2\pi i f t} dt
    \end{split}
\end{equation}
The spectrum is then given by
\begin{equation}
\label{eqn:spectrum_A}
    \begin{split}
        h(f) = 
        \frac{2GM}{Rc^{4}}& 
        \left(\frac{iA^{2}}{4\pi^{3}f^{3}}
        \left\{
        1 - e^{-2\pi i f T}
        \right\} + \right.\\
        & \frac{1}{2\pi^{2}f^{2}}
        \left\{
        A^{2}Te^{-2\pi i f T}
        -
        Av_{0}\left(1 - e^{-2\pi i f T}\right)
        \right\}\\
        & + \left.\frac{i}{2\pi f}
        \left\{
        A^{2}T^{2}e^{-2\pi i f t_{2}}
        +
        2ATv_{0}e^{-2\pi i f t_{2}}\right.
        \right.\\
        &
        \left.
        \left.
        +
        v_{0}^{2}
        \left(
        e^{-2\pi i f t_{2}}
        -
        e^{-2\pi i f t_{1}}
        \right)
        \right\}
        \right)
    \end{split}
\end{equation}
where the term proportional to $v_{0}^{2}$ is zero for discrete data where $f = k\tau^{-1}$ with $k\in \mathbb{Z}$. The spectrum (\ref{eqn:spectrum_A}) can be generalized to the other three cases by taking
\begin{enumerate}
    \item $T\to T$
    \item $T \to t_{2}$
    \item $T \to T - t_{1}$
    \item $T \to t_{2} - t_{1} = \tau$
\end{enumerate}
for each case. For future reference, we refer to these $T$ modifications as the `perceived acceleration period' of the data sample. 
\par
It is also instructive to view the spectrum $(\ref{eqn:spectrum_A})$ in terms of the change in velocity $\Delta v = AT$. With this substitution, the spectrum becomes
\begin{equation}
\label{eqn:spectrum_dv}
    \begin{split}
        h(f) = 
        &\frac{2GM}{Rc^{4}}
        \left(\frac{i}{4\pi^{3}f^{3}}
        \left(\frac{\Delta v}{T}\right)^{2}\left\{
        1 - e^{-2\pi i f T}
        \right\} + \right.\\
        & \frac{1}{2\pi^{2}f^{2}}
        \left\{
        \left(\frac{\Delta v^{2}}{T}\right)e^{-2\pi i f T}
        +\right.\\
        &\left.\left(\frac{\Delta v \cdot v_{0}}{T}\right)\left(1 - e^{-2\pi i f T}\right)
        \right\} +\\
        & \frac{i}{2\pi f}
        \left\{
        \Delta v^{2}e^{-2\pi i f t_{2}}
        +
        \Delta vv_{0}e^{-2\pi i f t_{2}}\right.\\
        &\left.\left.
        +
        v_{0}^{2}e^{-2\pi i f t_{1}}
        \left(
        e^{-2\pi i f (t_{2}-t_{1})}
        -
        1
        \right)
        \right\}
        \right)
    \end{split}
\end{equation}
The form (\ref{eqn:spectrum_dv}) is useful because the $f^{-1}$ term is independent of $T$. As we show next, the spectrum is dominated by this term, so the dependency of the spectrum on $T$ is largely mitigated using (\ref{eqn:spectrum_dv}) (see Appendix \ref{app:b}).
\subsubsection{Frequency-Scaling Behavior}\hfill \\
\vspace{-0.4cm}
\par 
The form of the spectrum (\ref{eqn:spectrum_A}) is encouraging because it suggests that objects with constant acceleration generate particularly strong signals for lower frequencies. This points to the use of  low-frequency, space-based detectors as a particularly promising avenue for the detection of such objects. Here we parse out exactly how the spectrum scales with frequency.
\par 
For simplicity, we consider Case \ref{case1} when discussing the frequency-scaling behavior. Furthermore, we assume that roughly the same time duration of both initial and final velocity contributions are present in the data sample. There are some slight variations when these assumptions are relaxed, but for the purposes of this study, this consideration is comprehensive. 
\par
\begin{figure}
    \centering
    \includegraphics[width = 1.0\linewidth]{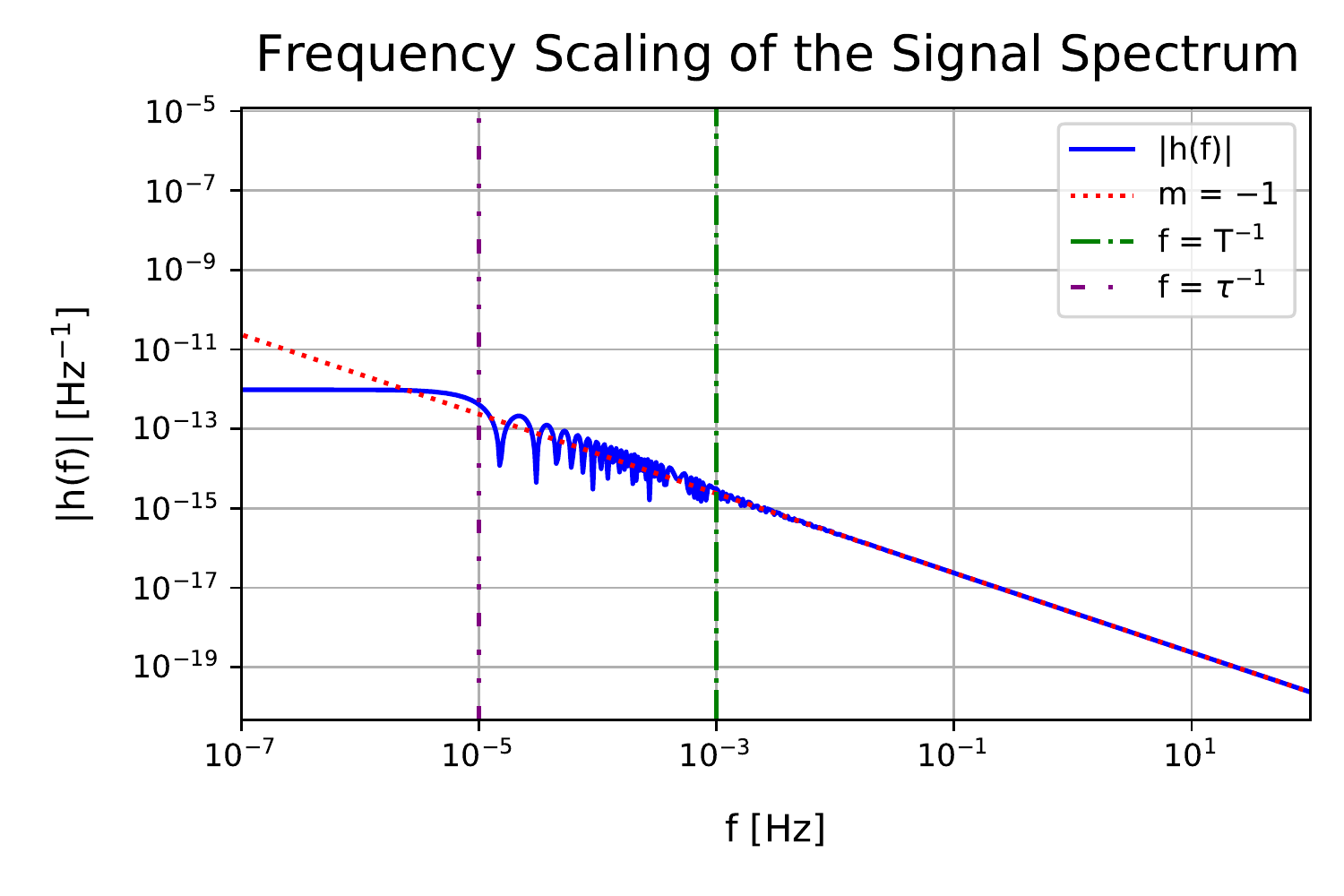}
    \caption{Plot of the GW spectrum (\ref{eqn:spectrum_A}) for an acceleration period of $T = $ 1,000\,s and a data sample length of $\tau = 10^{5}$\,s. We can see that the spectrum continues to scale like $f^{-1}$ past $f^{-1} = T^{-1}$ until $f = \tau^{-1}$,  after which the spectrum levels out.}
    \label{fig:2}
\end{figure}

 
We ultimately find that the spectrum scales like $f^{-1}$ for frequencies $f > \tau^{-1}$ and $f^{0}$ for frequencies $f < \tau^{-1}$, which are irrelevant for discrete data (see Figure \ref{fig:2}). To illustrate why this is the case, we start by showing that the dynamic contribution $I_{2}$ scales like $f^{-1}$ for frequencies $f > T^{-1}$ and $f^{0}$ for $f < T^{-1}$, even for $T > 1$\,s. This result is somewhat unintuitive because the dynamic contribution $I_{2}$ contains $f^{-2}$ and $f^{-3}$ terms. However, for frequencies $f > T^{-1}$, the following inequality is necessarily satisfied:
\begin{equation}
\label{eqn:T1_condition}
    \frac{T^{n-1}}{f}
    \geq 
    \frac{1}{f^{n}}
\end{equation}
Then, since each $f^{-2}$ and $f^{-3}$ term has a corresponding $f^{-1}$ term with a relative factor of $T$ and $T^{2}$, respectively, the $f^{-1}$ behavior will dominate for $f > T^{-1}$. For lower frequencies, the $I_{2}$ contribution plateaus because the phase in (\ref{eqn:fourier_integral}) becomes much larger than the acceleration period $T$. Therefore, the integrands become relatively constant as the frequency decreases further, and the spectrum plateaus.
\par 
In the case where only the dynamic portion of the signal is captured $\tau < T$ (Case \ref{case4}), the effect is to change the `perceived acceleration period' from $T \to \tau$, and the previous argument holds for frequencies $f > \tau^{-1}$. 
\par 
In the case where $\tau > T$ and both constant velocity contributions are captured with roughly equal length, the contributions from both $I_{1}$ and $I_{3}$ scale like $f^{-1}$ until the integrands begin to saturate at $f < \tau^{-1}$. Then, since the dynamic contribution $I_{2}$ plateaus at $f < T^{-1}$ and $\tau > T$ by assumption, the $f^{-1}$ contributions from $I_{1}$ and $I_{3}$ will dominate for $T^{-1} < f < \tau^{-1}$. So to summarize, the spectrum scales like $f^{-1}$ for $f > \tau^{-1}$, which is the entire frequency domain for discrete data. 
\par 
In fact, we can show that any sufficiently smooth linear trajectory will scale like $f^{-1}$ for $f \gtrsim \tau^{-1}$ if the trajectory does not undergo rapid oscillations during the acceleration period. If we remain agnostic about the form of the velocity during the acceleration duration and Taylor-expand on the interval $[0,T]$,
\begin{equation}
\label{eqn:taylor}
    v(t) = 
    v_{0} + v_{1}\left(\frac{t}{T}\right) + v_{2}\left(\frac{t}{T}\right)^{2} + v_{3}\left(\frac{t}{T}\right)^{3} + ...
\end{equation}
we can again compute the dynamic contribution to the spectrum trajectory:
\begin{equation}
    I_{2}
    =
  \int_{0}^{T}v^{2}(t)e^{-2\pi i f t}dt  
\end{equation}
Since the higher power contributions $f^{-n}$ come only from $I_{2}$, we find that each of these terms has the form
\begin{equation}
    h(f^{-n})
    \propto
    \frac{v_{i}v_{j}e^{-2\pi i fT}}{\pi^{n}f^{n}T^{n-1}}
\end{equation}
where $v_{i}$ is the $i$th coefficient in the expansion (\ref{eqn:taylor}). The spectrum then clearly scales like $f^{-1}$ for $f > T^{-1}$ by (\ref{eqn:T1_condition}).  
\par Then for lower frequencies, we still have the constant velocity contributions that scale like $f^{-1}$. Before, we knew that these contributions would dominate for $f < T^{-1}$ because the dynamic contribution plateaus for these frequencies. This need not be the case for a general trajectory, particularly if $v(t)$ oscillates rapidly on the interval $[0, T]$. However, for any sufficiently smooth trajectory, the previous arguments will hold, and the spectrum will scale like $f^{-1}$ for $f \gtrsim \tau^{-1}$ \footnote{We use $\gtrsim$ as opposed to $>$ to allow for small variations due to oscillations in $v(t)$}. 
\par
\begin{figure}
\centering
\includegraphics[width=1\linewidth]{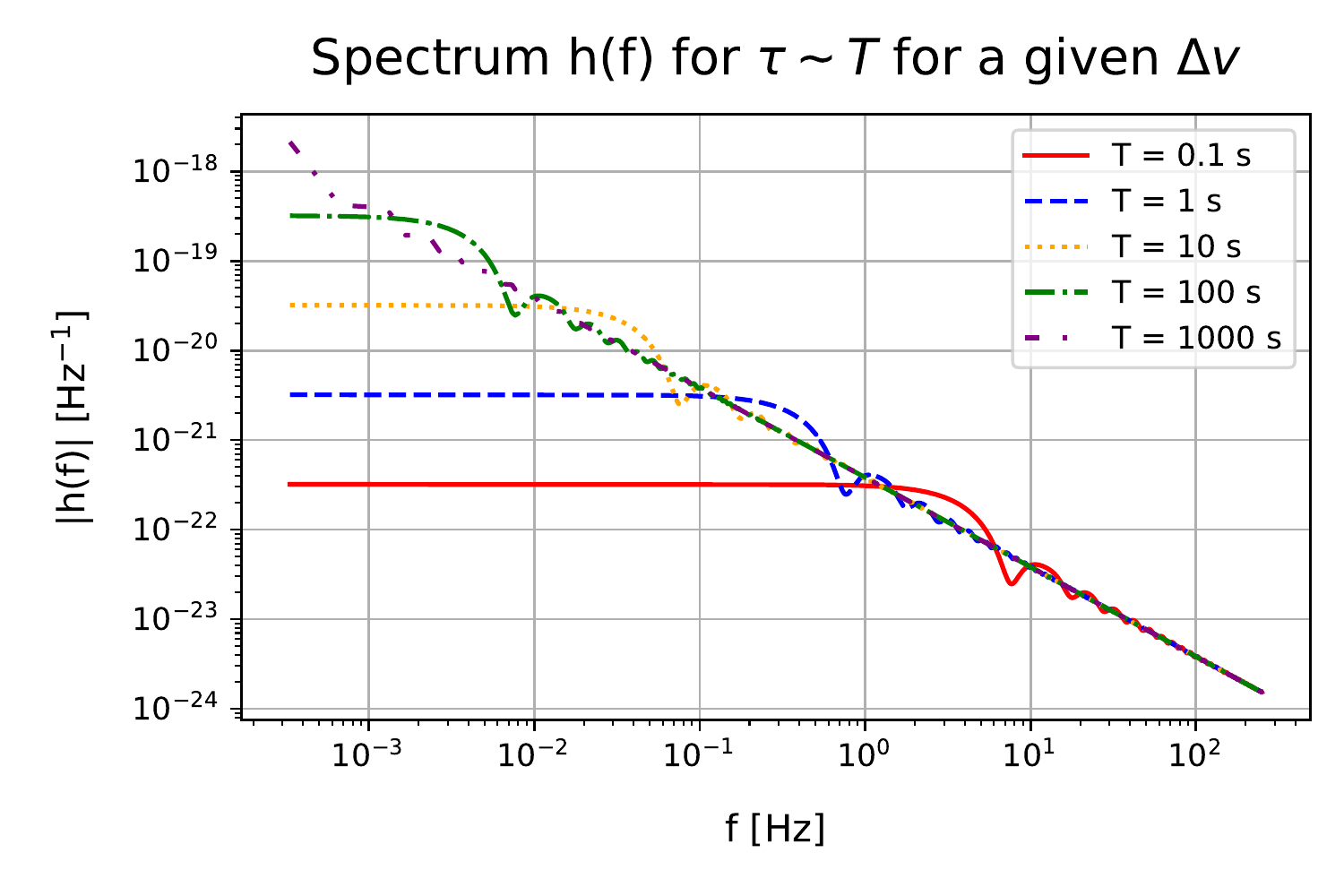}
\includegraphics[width=1\linewidth]{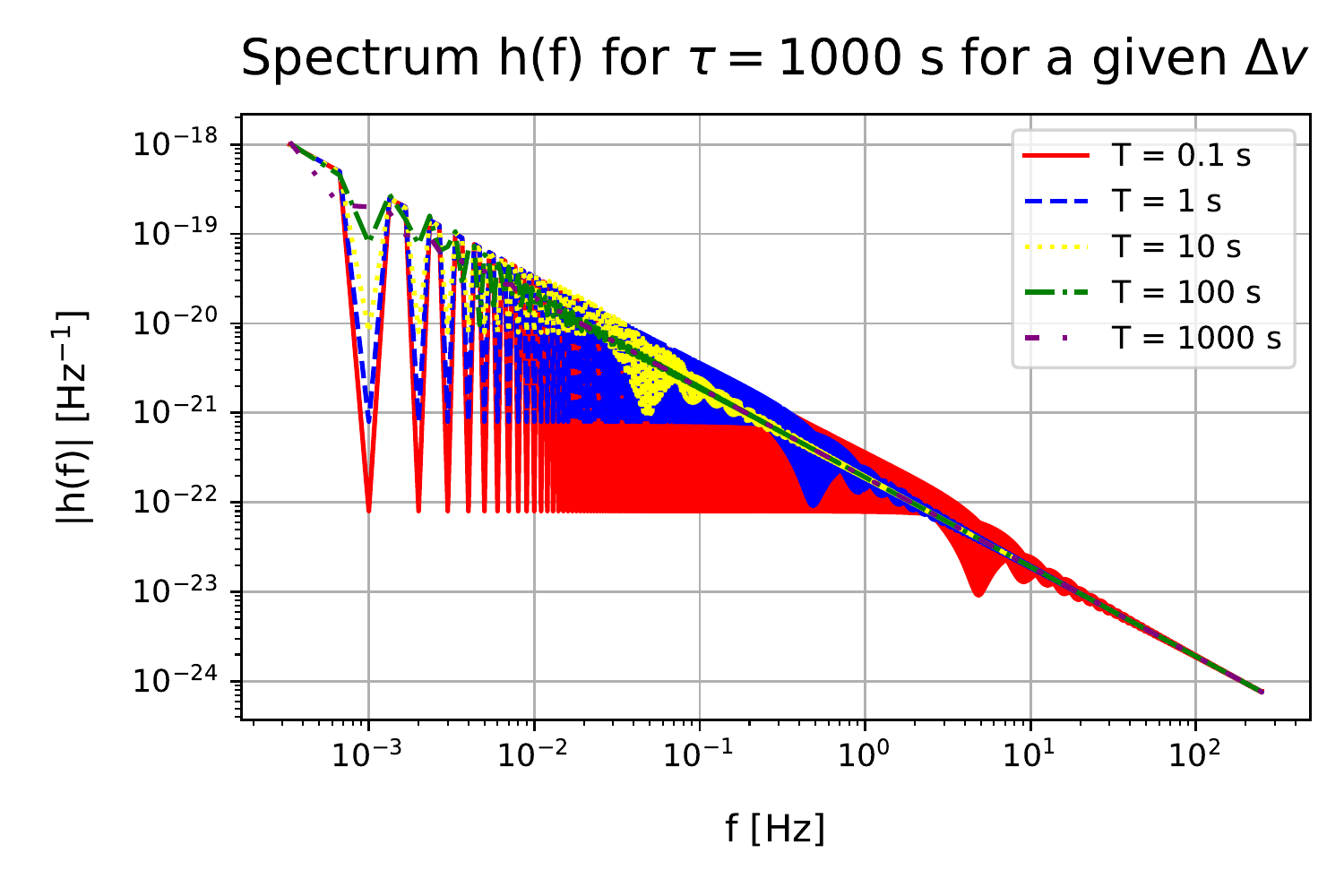}
\caption{Plot of spectrum for a fixed change in velocity (\ref{eqn:spectrum_dv}) for varying acceleration periods $T$ for two cases, namely where the data sample length is made comparable to the acceleration period $\tau = 3T$ (Top) and where $\tau$ is fixed at the maximum value of $T$ plotted ($T =$ 1,000$\,{\rm s}$) (Bottom). We choose to plot these values for a given $\Delta v$ because the $f^{-1}$ behavior is approximately independent of $T$ in this case (see (\ref{eqn:spectrum_dv})). We see that for $\tau \sim T$, the spectrum plateaus at $f \sim T^{-1}$, whereas the $f^{-1}$ behavior is extended past these plateau points when $\tau$ is fixed to a larger value, up to some oscillations.}
\label{fig:3}
\end{figure}

 
The significance of this scaling behavior is two-fold. First, since the spectrum scales like $f^{-1}$, the signal strength is clearly stronger for lower frequencies, so detectors that are more sensitive to lower frequencies will more easily detect these objects (for details, see Section \ref{sec:Discussion}). Second, for an object that accelerates for a period $T$, the $f^{-1}$ behavior can be seen for frequencies $f < T^{-1}$ by extending the data sample duration $\tau > T$, thereby capturing the gravitational memory of the initial and final velocity. This effect can be seen in Figure \ref{fig:3}. As we discuss in more detail in Section \ref{sec:Discussion}, this is significant because it implies that low-frequency detectors will not only be sensitive to objects that accelerate for periods greater than or equal to the characteristic time scale of the detector, but for shorter times as well. Granted, this effect is contingent on the assumption (\ref{eqn:const_R}) not being violated as $\tau$ is increased, but as we show in Section \ref{sec:Discussion}, this is not a particularly stringent requirement for objects at parsec-scale distances from Earth.
\section{Detecting Constant Acceleration}
\label{sec:Results}
We now move on to computing the distance at which LIGO and LISA can detect \textsc{RAMAcraft} with constant acceleration.
\subsection{Detection Statistics} Before we determine detection ranges for these signals, we first review the detection statistics from the literature that we will use to quantify the strength of GW signals. Specifically, we discuss the search methods commonly used at detector collaborations, namely transient burst and matched-filtering (MF) searches. But first, we review detector sensitivity curves.

\subsubsection{Detector Sensitivity Curves}\hfill \\
When a GW passes through an object, it causes it to expand and contract. Detectors then register a signal (or strain) $h$ by tracking these movement patterns. The most well-established of these detectors are massive interferometers such as LIGO, Virgo, and KAGRA. Using kilometer-scale lasers, these interferometers can detect displacements down to one ten-thousandth of the width of a proton \citep{facts_caltech}.
\par
In principle, these detectors can detect strains $h$ of any size. However, as with any detector, this prospect is limited by the noise profile. The noise profile of these detectors has been carefully scrutinized for many years for the purposes of noise mitigation \citep{Abbott_2020_noise}. Such noise mitigation efforts have proven to be a demanding engineering challenge; the detectors are so sensitive that distant airplanes, passing trucks, and even tumbleweeds can cause interference \citep{Berger_2004}. 
\par
\begin{figure}
\centering
\includegraphics[width=1.0\linewidth]{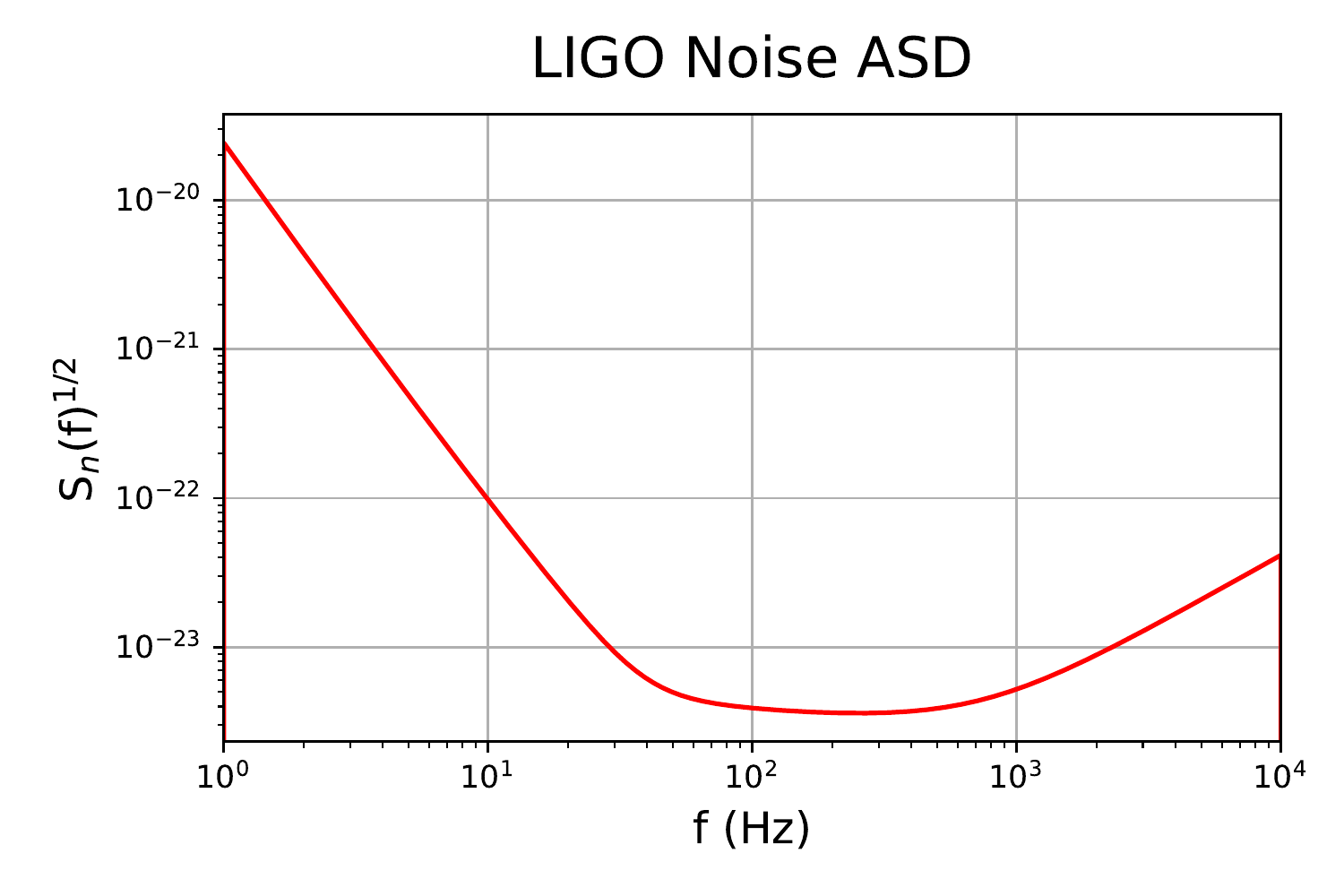}
\includegraphics[width=1.0\linewidth]{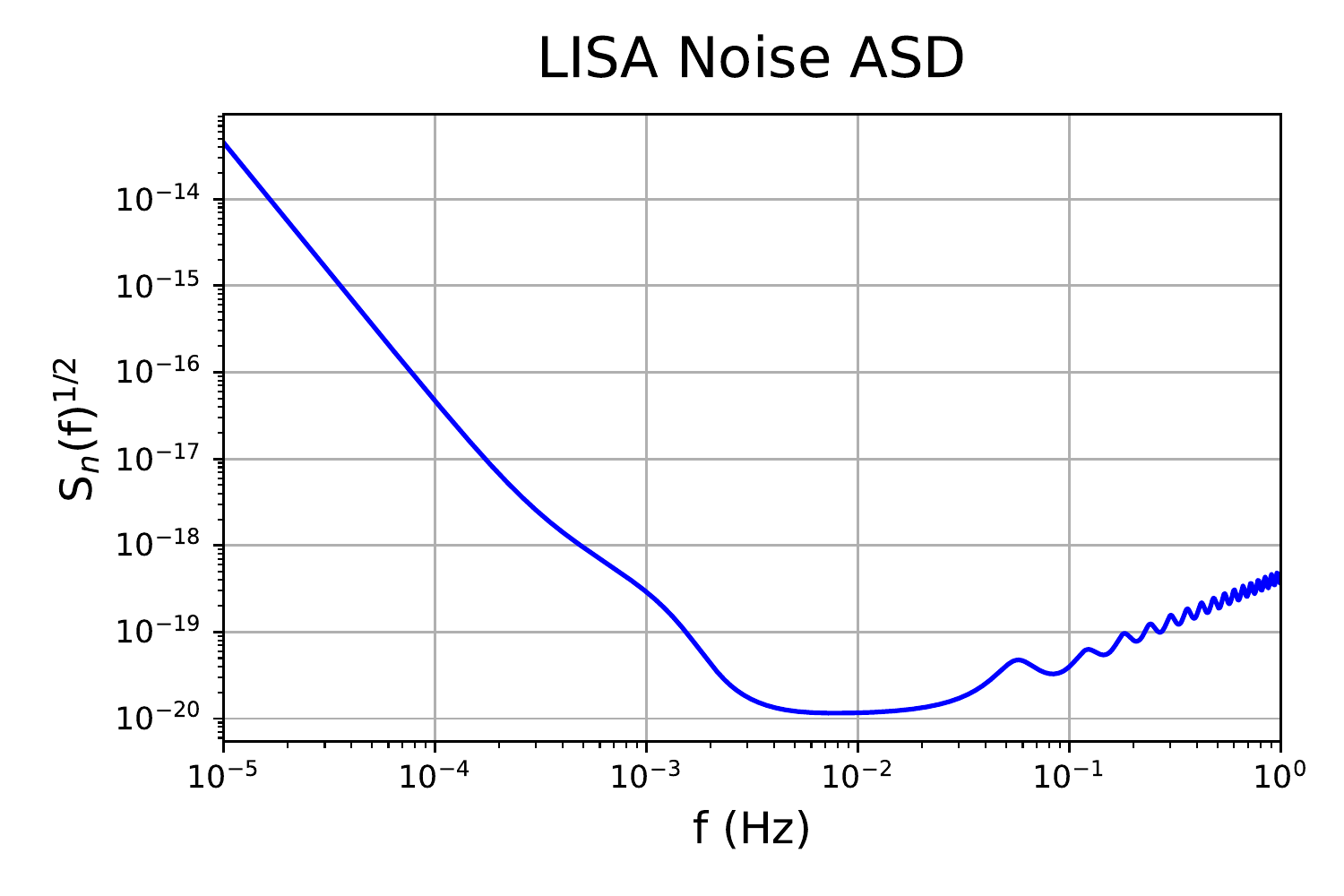}
\caption{(Top) Plot of LIGO sensitivity curve generated using the aLIGOZeroDetHighPower method from the PyCBC library \citep{pycbcsw}. (Bottom) LISA sensitivity curve taken from the GitHub provided in \citet{LISA_sens}.}
\label{fig:4}
\end{figure}

Since the noise profile sets a detection threshold, it also determines the sensitivity of the detector. For this reason, the terms `noise profile' and `sensitivity curve' are interchangeable.
The sensitivity curve of LIGO \citep{Abbott_2020_noise} and the projected sensitivity curve of LISA \citep{LISA_sens} can be seen in Figure \ref{fig:4}. Specifically, the curves in Figure \ref{fig:4} are amplitude spectral densities (ASDs) averaged over many data samples, giving them a colored Gaussian noise profile. For real-time detection, stochastic glitches also contribute to the noise $n(t)$, so the colored Gaussian noise description is an approximation. 
\par
The ASD is the square root of the power spectral density (PSD), which in the continuous case is given by
\begin{equation}
    \delta(f-f')S_{n}(f)
    =
    \langle
    n(f) \tilde{n}(f')
    \rangle
\end{equation}
$S_{n}(f)$ is essentially an ensemble average of $|n(f)|^{2}$ where $n(f)$ is a CFT of a pure noise input. However, in the discrete case, a different convention is typically used depending on the convention used to take the FFT of time-domain samples. In our case, we use the convention from \citet{find_chirp}, which matches the output of the CFT of a time-domain signal with compact support within the sample duration $\tau$. Then if $n(f_{k})$ is an FFT generated using this convention, we find that the curves in Figure \ref{fig:4} and $n(f_{k})$ are related by (see Figure \ref{fig:5})
\begin{equation}
    S_{n}(f_{k})
    =
    \langle
    |n(f_{k})|^{2}
    \rangle
    \cdot 
    (2\Delta f)
\end{equation}
where $\Delta f$ is the frequency-grid spacing. The reason for the $2\Delta f$ factor is two-fold. The factor of 2 accounts for the fact that the sensitivity curves $S_{n}$ in Figure 4 are one-sided PSDs, and the factor of $\Delta f = \tau^{-1}$ accounts for the fact that the noise FFTs scale like $n(f_{k})\propto \tau^{1/2}$. 
\subsubsection{Transient Burst Searches}
One straightforward way to deduce whether or not a signal is detectable is to determine whether or not the signal rises above the detector sensitivity curve.
This is essentially the idea behind burst searches, which search for excess power within some frequency band. This category of searches does not impose any assumptions on the shape of the signal, so these search methods apply to all possible GW shapes.
\par
The well-established burst detection pipelines used, for example, in the burst searches \citep{burst1, burst2, string_burst} are the Coherent Wave Burst (cWB) \citep{cWB} and the Omicron-LIB (oLIB) \citep{omicron} pipelines. The statistic used in both of these pipelines is the \textit{excess power} statistic. For a continuous signal with spectrum $h(f)$, the excess power within some frequency band is given by
\begin{equation}
\label{eqn:pb_int}
    \rho_{\rm b}^{2} = 4\int^{f_{2}}_{f_1}{\frac{|h(f)|^{2}}{S_{n}(f)} df}
\end{equation}
where $S_{n}(f)$ is again the power spectral density (PSD) of the detector and $f_{1}$ and $f_{2}$ determine the choice of frequency band. For discrete data $h(f_{k})$, the same statistic can be implemented as in \citet{discrete_burst}:
\begin{equation}
\label{eqn:pb_sum}
\rho_{\rm b}^{2} = 4 \sum_{k_{1}}^{k_{2}}{\frac{|h(f_{k})|^{2}}{S_{n}(f_{k})}\Delta f}
\end{equation}
where $\Delta f = \tau^{-1}$ is the frequency resolution and $\tau$ is the data sample duration. In both cases, $S_{n}$ should be normalized such that the excess power output for a purely noise input is $\sigma_{\rm b}^{2} \approx 1$. The alternative is to compute the excess power for a pure noise input $n(f)$:
\begin{equation}
\label{eqn:sigb_int}
    \sigma_{\rm b}^{2} = 4\int^{f_{2}}_{f_1}{\frac{|n(f)|^{2}}{S_{n}(f)} df}
\end{equation}
and to find the signal-to-noise ratio (SNR) by taking $\rho_{\rm b}^{2}\to\rho_{\rm b}^{2}/\sigma_{\rm b}^{2} = (S / N)^{2}_{\rm b}$. Since the noise will contain stochastic glitches, this is not possible in general, but this can be done to a reasonable degree by normalizing the output from noise data generated via the sensitivity curves in Figure \ref{fig:4}, as we do later in our analysis. 
\subsubsection{Matched Filtering}
The burst SNR will only reach a detection threshold of $\rho^{2}_{\rm b} / \sigma^{2}_{\rm b} > 1$ for signals whose spectrum exceeds the noise in some frequency band. Ideally, we'd like to use a detection method that is valid for signals that are everywhere below the noise level. Such a method is provided by matched-filtering (MF).
\par
The optimal detection statistic for processing signals embedded in Gaussian noise is the MF SNR. Although interferometer noise contains glitches that are non-Gaussian, glitch mitigation in combination with MF \citep{find_chirp} has allowed GW pipelines to detect signals below the noise threshold \citep{below_noise_detection}. 
\par 
For a strain comprised of both a signal and a noise part $h(t) = s(t) + n(t)$, the MF statistic is given by \citet{find_chirp}
\begin{equation}
\label{eqn:pmf_int}
\rho_{\rm MF}^{2}
=
4\Bigg\lvert
\mathcal{R}\left\{
\int_{-\infty}^{\infty}{\frac{s(f) \tilde{h}(f)}{S_{n}(f)}}df
\right\}
\Bigg\rvert
\end{equation}
where $\tilde{h}(f)$ is the complex conjugate of the strain spectrum for the signal being searched for. In the discrete case, we have
\begin{equation}
\label{eqn:pmf_sum}
\rho_{\rm MF}^{2} = 4\Bigg\lvert 
\mathcal{R}\left\{
\sum_{k_{1}}^{k_{2}}{\frac{s(f_{k}) \tilde{h}(f_{k})}{S_{n}(f_{k})}\Delta f}
\right\}
\Bigg\rvert
\end{equation}
In the ideal case where the strain is comprised entirely of the signal, the output (\ref{eqn:pmf_int}) is the same as (\ref{eqn:pb_int}). However, the MF output for a pure noise signal:
\begin{equation}
\label{eqn:sigmf_int}
    \sigma_{\rm MF}^{2}
    =
    4
    \Bigg\lvert
    \mathcal{R}\left\{\int_{-\infty}^{\infty}{\frac{n(f) \tilde{h}(f)}{S_{n}(f)}}df
    \right\}
    \Bigg\rvert
\end{equation}
is typically much smaller than the excess-power output of pure noise (\ref{eqn:sigb_int}) because the noise and the signal spectrum should be relatively uncorrelated. Consequently, the discrepancy in statistic output $\rho^{2}$ for signals and noise is typically much larger for MF searches than for burst searches, making MF the more optimal method. 
\subsubsection{Modified Statistic and $\rho^{2}$ vs $\tau$ Scaling}\hfill \\
\par
\begin{figure}
    \centering
    \includegraphics[width = 1.0\linewidth]{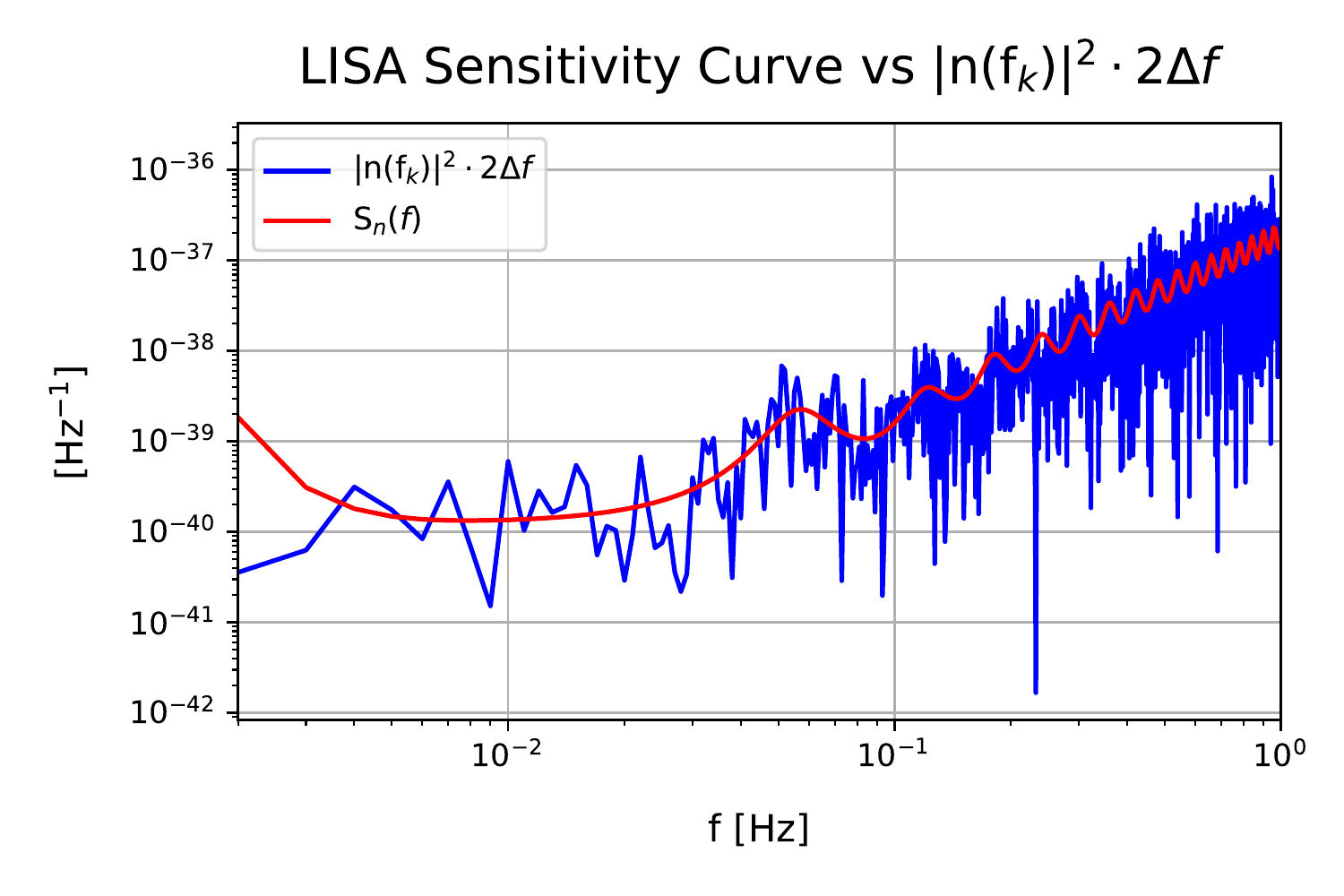}
    \caption{Plot of LISA sensitivity curve $S_{n}(f_{k})$ vs $|n(f_{k})|^{2}\cdot 2\Delta f$, where $n(f_{k})$ is an FFT of the time-domain noise data simulated from $S_{n}(f_{k})$ using the colored\_noise method from pycbc.noise.reproduceable. The ratio is very close to 1 throughout the frequency interval, yielding $\sigma^{2} \approx 1$ using the convention (\ref{eqn:noise_burst_SNR}).}
    \label{fig:5}
\end{figure}

 
We choose to implement the detection statistic such that the burst statistic $\sigma^{2}_{\rm b}$ for a pure noise input is approximately 1. This is achieved by (see Figure \ref{fig:5})
\begin{equation}
\label{eqn:pb}
    \rho_{\rm b}^{2}
    =
    \frac{2}{N}
    \sum_{k=1}^{N}
    {\frac{|h(f_{k})|^{2} \Delta f}{S_{n}(f_{k})}}
\end{equation}
for the signal output and
\begin{equation}
\label{eqn:noise_burst_SNR}
    \sigma_{\rm b}^{2}
    =
    \frac{2}{N}
    \sum_{k=1}^{N}
    {\frac{|n(f_{k})|^{2} \Delta f}{S_{n}(f_{k})}}
\end{equation}
for the noise output,
where $N$ is the number of discrete points within the sensitivity band of interest. Similarly, for MF searches we use
\begin{equation}
\label{eqn:pmf}
    \rho_{\rm MF}^{2}
    =
    \frac{2}{N}
    \sum_{k=1}^{N}
    {\frac{s(f_{k})\cdot\tilde{h}(f_{k})\Delta f}{S_{n}(f_{k})}}
\end{equation}
for the signal output and 
\begin{equation}
\label{eqn:sigmf}
    \sigma_{\rm MF}^{2}
    =
    \frac{2}{N}
    \sum_{k=1}^{N}
    {\frac{n(f_{k})\cdot\tilde{h}(f_{k})\Delta f}{S_{n}(f_{k})}}
\end{equation}
for the noise output.
We can make such a choice because the statistics $\rho^{2}$ and $\sigma^{2}$ differ from  the conventions (\ref{eqn:pb_int} -- \ref{eqn:sigmf_int}) by the same factor $(2N)^{-1}$, and their ratio is what yields the SNR.
\par 
With the discrete statistics (\ref{eqn:pb}) and (\ref{eqn:noise_burst_SNR}) defined, we conclude here by explicitly clarifying the scaling of each of these statistics with respect to the data sample length $\tau$. 
\par 
The scaling for $\rho_{\rm b} / \sigma_{\rm b}$ is fairly straightforward. As we mentioned earlier, we find that the noise spectrum scales like $n(f_{k}) \propto \tau^{1/2}$. From the factor of $\Delta f$ in (\ref{eqn:noise_burst_SNR}), we find $\sigma_{\rm b}^{2} \propto \tau^{0}$. Then assuming that our signal corresponds to Case \ref{case1}, the spectrum strength does not depend on the data sample length $h(f_{k})\propto\tau^{0}$, so $\rho^{2}_{\rm b} \propto \tau^{-1}$, and the burst SNR scales like
\begin{equation}
\label{eqn:snrb_tau}
    \frac{\rho_{\rm b}}{\sigma_{\rm b}}
    =
    \left(\frac{S}{N}
    \right)_{\rm b}
    \propto
    \tau^{-1/2}
\end{equation}
The scaling behavior for MF searches is slightly more complicated because there is an extra effect that cannot be read from the definitions (\ref{eqn:pmf} - \ref{eqn:sigmf}). Using just these definitions and carrying out a similar analysis to (\ref{eqn:snrb_tau}), we would find that $\sigma_{\rm MF} \propto \tau^{-1/4}$ and $\rho_{\rm MF} \propto \tau^{-1/2}$. However, because the noise correlation $\sigma_{\rm MF}$ is computed as a discrete sum, the output depends on how many points are used. Intuitively, if one samples a sine wave, the odds that the sum of two sampled points is very small is quite low, but the odds increase as many points are taken over many wavelengths. Empirically, we find that this effect contributes an extra scaling of $\sigma_{\rm MF}\propto \tau^{-1/4}$ (see Appendix \ref{app:d}). Therefore, we find that the MF SNR scales like
\begin{equation}
\label{eqn:MF_SNR_tau}
    \frac{\rho_{\text{\rm MF}}}{\sigma_{\text{\rm MF}}}
    =
    \left(\frac{S}{N}
    \right)_{\rm MF}
    \propto 
    \tau^{0}
\end{equation}
Both of these results (\ref{eqn:snrb_tau}) and (\ref{eqn:MF_SNR_tau}) are for an object that satisfies Case \ref{case1} from the previous section. For an object where the data sample ends before the acceleration ceases, the spectrum (\ref{eqn:spectrum_dv}) is modified to the perceived change in velocity: $\Delta v = AT \to \Delta v' = At_{2}$. Then, as $\tau$ increases, the perceived change in velocity increases proportionally to $\tau$, and since the $f^{-1}$ part of the spectrum scales like $h(f^{-1})\propto \Delta v'^{2}$ for small $v_{0}$, we find that
\begin{equation}
\label{SNRb_tau_bigT}
    \frac{\rho_{\rm b}}{\sigma_{\rm b}}
    =
    \left(\frac{S}{N}
    \right)_{b}
    \propto
    \tau^{3/2}
\end{equation}
and 
\begin{equation}
\label{eqn:MF_SNR_tau_bigT}
    \frac{\rho_{\text{\rm MF}}}{\sigma_{\text{\rm MF}}}
    =
    \left(\frac{S}{N}
    \right)_{\rm MF}
    \propto 
    \tau^{1}
\end{equation}
for $t_{2} < T$ and small $v_{0}$.  These results will be important to consider in the next section where we present the ranges for both LISA and LIGO using various data sample lengths. 
\subsection{Detection Analysis}
We are now prepared to compute the ranges at which LIGO and LISA can detect linearly-moving \textsc{RAMAcraft} undergoing constant acceleration. 
Specifically, we will compute detection ranges for both burst and MF searches to account for the two cases for which these signals are either passively (burst) or actively (MF) searched. For simplicity, we compute both statistics for the ideal case where the strain $h$ is devoid of noise $s(t) = h(t)$. In this case, the two raw outputs (\ref{eqn:pb}) and (\ref{eqn:pmf}) are equivalent. We then differentiate the burst SNR from the MF SNR by applying the appropriate normalizations (\ref{eqn:noise_burst_SNR}) and (\ref{eqn:sigmf}), respectively. Then choosing a detection threshold of $(S / N)_{\rm det} = 8$, we determine the detectability of linearly-accelerating trajectories with small initial velocity $v_{0}$ based on their mass $M$, change in velocity $\Delta v$, and distance from the detector $R$. We use the change in velocity $\Delta v$ as opposed to the acceleration $A$ because the spectrum is effectively independent of the acceleration duration $T$ for a given $\Delta v$, thereby eliminating a variable from our analysis. We choose signals with small initial velocity $v_{0}$ both because the results that we find apply to objects with large changes in velocity, and also because setting $v_{0} \to 0$ allows us to express the range $R$ in terms of one variable $M\Delta v^{2}$ (see (\ref{eqn:no_v0_spectrum})).
\par
Furthermore, since the burst detection pipelines are indifferent to signal shape, linearly-accelerating signals that fall within the detectable regime for LIGO burst searches should have already been detected (if the signal occurred while LIGO was online). The lack of burst detections thereby constrains the parameter space for linearly-accelerating trajectories. 
\subsubsection{Methodology}\hfill \\
For this analysis, we use discrete data and data analysis packages from both the NumPy and PyCBC \citep{pycbcsw} libraries. The key steps involved for a given parameter set $\{M, \Delta v, T, R\}$, detector (LIGO or LISA), and detection method (burst or MF) are the following:
\begin{enumerate}
    \item Choosing some test parameter set $\left\{M, \Delta v, T, R_{0}\right\}$, generate a time-domain signal $h(t_{n})$ according to (\ref{eqn:time_strain}).
    \item Generate a frequency-domain signal $h(f_{k})$ using the convention (\ref{eqn:fft_convention}).
    \item Import data for either the LIGO or LISA noise sensitivity curve $S_{n}(f_{k})$. 
    \item Simulate time-domain noise data $n(t_{n})$ using the sensitivity curve.
    \item Generate a frequency-domain noise signal $n(f_{k})$ using the convention (\ref{eqn:fft_convention}).
    \item Depending on the search method, compute $\sigma^{2}$ for the noise input $n(f_{k})$ using either (\ref{eqn:noise_burst_SNR}) or (\ref{eqn:sigmf}). 
    \item Depending on the search method, compute the detection statistic $\rho^{2}$ by either (\ref{eqn:pb}) or (\ref{eqn:pmf}).
    \item `Divide out' the test parameter set by calculating a `reduced SNR' $(S / N)_{\rm red}$ for either a burst or MF search. This allows us to factor out the parameter dependence of the SNR and solve for $R$ for a given $M$ and $\Delta v$.
    \item Choose a detection threshold $(S / N)_{\rm det}$.
    \item With the calculated `reduced SNR' and the parameter dependence factored out, solve for $R$ for a given $M\Delta v^{2}$.
\end{enumerate}
\par
Here we go over these steps in more detail where necessary. 
To generate the data $h(f_{k})$, we use the numpy.fft library, which uses the convention (\ref{eqn:fft_convention}) out of the box. To import data for the noise sensitivity curves $S_{n}(f_{k})$, we use the aLIGOZeroDetHighPower method from the pycbc.psd library for the case of LIGO. For LISA, we use the sensitivity curve data provided in the GitHub from \citet{LISA_sens} and interpolate the data on the frequency grid of $h(f_{k})$.
To simulate noise $n(t_{n})$ using the sensitivity curves $S_{n}(f_{k})$, we use the colored\_noise method from pycbc.noise.reproduceable. 
\par 
The sensitivity curve constitutes an ensemble average of many noise samples, which is why we need to generate a simulated noise spectrum $n(f_{k})$ when we already have a sensitivity curve $S_{n}(f_{k})$. Simulating a noise sample $n(t_{n})$ and taking its spectrum $n(f_{k})$ better represents what one noise data sample will look like, and therefore better represents the output $\sigma^{2}$ if a pure noise data sample is collected. 
\par 
Here we explain how to compute the range $R$ after computing statistics for a test parameter set. By taking $v_{0} \to 0$, our spectrum for a given change in velocity $\Delta v$ (\ref{eqn:spectrum_dv}) becomes
\begin{equation}
\label{eqn:no_v0_spectrum}
h(f)
=
    \frac{2GM\Delta v^{2}}{Rc^{4}}
    \left\{
    \frac{ie^{-2\pi i ft_{2}}}{2\pi f}
    +
    \frac{e^{-2\pi i fT}}{2\pi^{2} f^{2}T}\right.
    \end{equation}
    \[
    \left. +
    \frac{i}{4\pi^{3} f^{3}T^{2}}
    \left(
    1 - e^{-2\pi i fT}
    \right)
    \right\}
    \]
Then if we factor out the parameter dependence of each spectrum by:
\begin{equation}
    h_{red}(f)
    =
    \left(
    \frac{2GM\Delta v^{2}}{Rc^{4}}
    \right)^{-1}
    h(f)
\end{equation}
we can rewrite the SNR quantities as 
\begin{equation}
\begin{split}
    \frac{\rho^{2}_{\rm b}}{\sigma^{2}_{\rm b}}
    =
    &\left(\frac{2GM\Delta v^{2}}{Rc^{4}}\right)^{2}
    \frac{\frac{2}{N}
    \sum_{k=1}^{N}
    {\frac{|h_{red}(f_{k})|^{2}\Delta f}{S_{n}(f_{k})}}}{\frac{2}{N}
    \sum_{k=1}^{N}
    {\frac{|n(f_{k})|^{2}\Delta f}{S_{n}(f_{k})}}}\\
    &=
    \left(\frac{2GM\Delta v^{2}}{Rc^{4}}\right)^{2}
    \left(\frac{S}{N}\right)^{2}_{\rm red, b}
\end{split}
\end{equation}
for burst searches and
\begin{equation}
\begin{split}
    \frac{\rho^{2}_{\rm MF}}{\sigma^{2}_{\rm MF}}
    =
    &\left(\frac{2GM\Delta v^{2}}{Rc^{4}}\right)
    \frac{\frac{2}{N}
    \sum_{k=1}^{N}
    {\frac{|h_{red}(f_{k})|^{2}\Delta f}{S_{n}(f_{k})}}}{\frac{2}{N}
    \lvert\mathcal{R}\left\{\sum_{k=1}^{N}
    {\frac{n(f_{k})\cdot\tilde{h}_{red}(f_{k})\Delta f}{S_{n}(f_{k})}}\right\}\rvert}\\
    &=
    \left(\frac{2GM\Delta v^{2}}{Rc^{4}}\right)
    \left(\frac{S}{N}\right)^{2}_{\rm red, MF}
\end{split}
\end{equation}
for MF searches. Then after computing the statistics $\rho^{2}$ and $\sigma^{2}$, we can calculate $(S / N)_{\rm red}$ by dividing out the test parameter set and solving for the range $R$ for a given $M\Delta v^{2}$ from the equation
\begin{equation}
    \left(\frac{S}{N}\right)^{2}_{\rm det}
    =
    \frac{\rho^{2}_{\rm b}}{\sigma^{2}_{\rm b}}
    =
    \left(\frac{2GM\Delta v^{2}}{Rc^{4}}\right)^{2}
    \left(\frac{S}{N}\right)^{2}_{\rm red, b}
\end{equation}
for burst searches and 
\begin{equation}
    \left(\frac{S}{N}\right)^{2}_{\rm det}
    =
    \frac{\rho^{2}_{\rm MF}}{\sigma^{2}_{\rm MF}}
    =
    \left(\frac{2GM\Delta v^{2}}{Rc^{4}}\right)
    \left(\frac{S}{N}\right)^{2}_{\rm red, MF}
\end{equation}
for MF searches. In particular, we see that $R \propto (S / N)_{\rm b}$ or $R \propto (S / N)^{2}_{\rm MF}$, depending on the search method.
\subsubsection{Results}\hfill 
\vspace{-0.5cm}
\par
Following the prescription just outlined, here we look at the ranges that an object producing the GW signal (\ref{eqn:time_strain}) would be detectable by LIGO or LISA. In particular, we take the limit $v_{0}\to 0$, which corresponds to objects whose change in velocity is much larger than their initial velocity (or vice versa if $A < 0$). In this case, we can plot the range $R$ as a linear function of one parameter $M\Delta v^{2}$, which gives the total change in kinetic energy of the \textsc{RAMAcraft} if its mass is approximately constant.
\par 
Before presenting the results, we explain here our choice of data sample parameters $\tau$, $t_{1}$, and $t_{2}$. For simplicity, we limit our considerations for now to Case \ref{case1}, where the entire acceleration duration is contained within the data sample. Specifically, we choose $t_{1} = -\tau / 3$ and $t_{2} = 2\tau / 3$ for a given $\tau$ in order to limit our study to Case \ref{case1}, which we do for the sake of consistency.
\par 
Our choice of data sample length $\tau$ then depends on the detection method. For burst searches, the SNR scales like $\tau^{-1/2}$, so we choose a $\tau$ not much smaller than that which makes the sensitivity band accessible. For LISA, we choose $\tau = 600$ s, and for LIGO, we choose $\tau = 0.1$ s. For MF searches, the SNR is independent of $\tau$, up to some oscillations to within a factor of a few (see Appendix \ref{app:d}). We then choose $\tau = 1,200$ s for LISA and $\tau = 300$ s for LIGO. 
\par 
The frequency band that we compute the statistics over also depends on the detection method. For burst searches, the SNR is essentially just an average of the ratio $h(f) / n(f)$, so including points only very close to the maximum value will result in the largest output. This is the idea behind band-pass burst searches \citep{FilterTechniques}. For our purposes, this can be achieved by sampling a small number of points about the sensitivity band minimum. However, it is necessary to sample enough points such that the noise statistic $\sigma^{2}$ is close to 1. This is unlikely to be the case when sampling just one or two points because the quantity $|n(f)|^{2} \Delta f$ oscillates about $S_{n}(f)$ (see Figure \ref{fig:5}). Empirically, we find that sampling 5 points about 1\,mHz for the case of LISA and 20 points about 100\,Hz for the case of LIGO (and the aforementioned frequency spacings $\tau^{-1}$) achieves this goal. 
\par 
For MF searches, one generally wants a much wider frequency range in order to achieve a small noise correlation (\ref{eqn:sigmf}). Empirically, we find not much difference in output when using a wide-band of $[10^{-3}, 1]$\,Hz for LISA and [1, 200] Hz for LIGO versus a moderate band of $[10^{-3}, 10^{-2}]$\,Hz for LISA and [10, 200] Hz for LIGO, so we choose the latter to conserve computational resources \footnote{Though the LIGO band extends into the kHz regime, since the signal scales like $f^{-1}$, we find not much benefit to integrating much further past 200 Hz.}.
\par 
To conclude our review of parameter choices, the choice of acceleration period $T$ from the test parameter set $\left\{M, \Delta v, T, R_{0}\right\}$ should not affect the results much due to the approximate independence of the spectrum on $T$ for a given $\Delta v$ (see Figure \ref{fig:b1}). Still, there are some slight variations due to the oscillations occurring for $f < T^{-1}$ (see Figure \ref{fig:3}). However, we find that the contributions of these oscillations is to cause variations in the SNR to within a factor of a few. We therefore just quote our choices of $T$ in generating the signals, namely $T = 10$\,s and 10\,ms for the LISA and LIGO burst search, respectively, and $T = 100$\,s and 10\,s for the LISA and LIGO MF search, respectively. The various acceleration periods $T$ were chosen such that $T \leq \tau$ for each search, ensuring that each test parameter set satisfied Case \ref{case1}. 
\par
\begin{figure}
\centering
\includegraphics[width=1.0\linewidth]{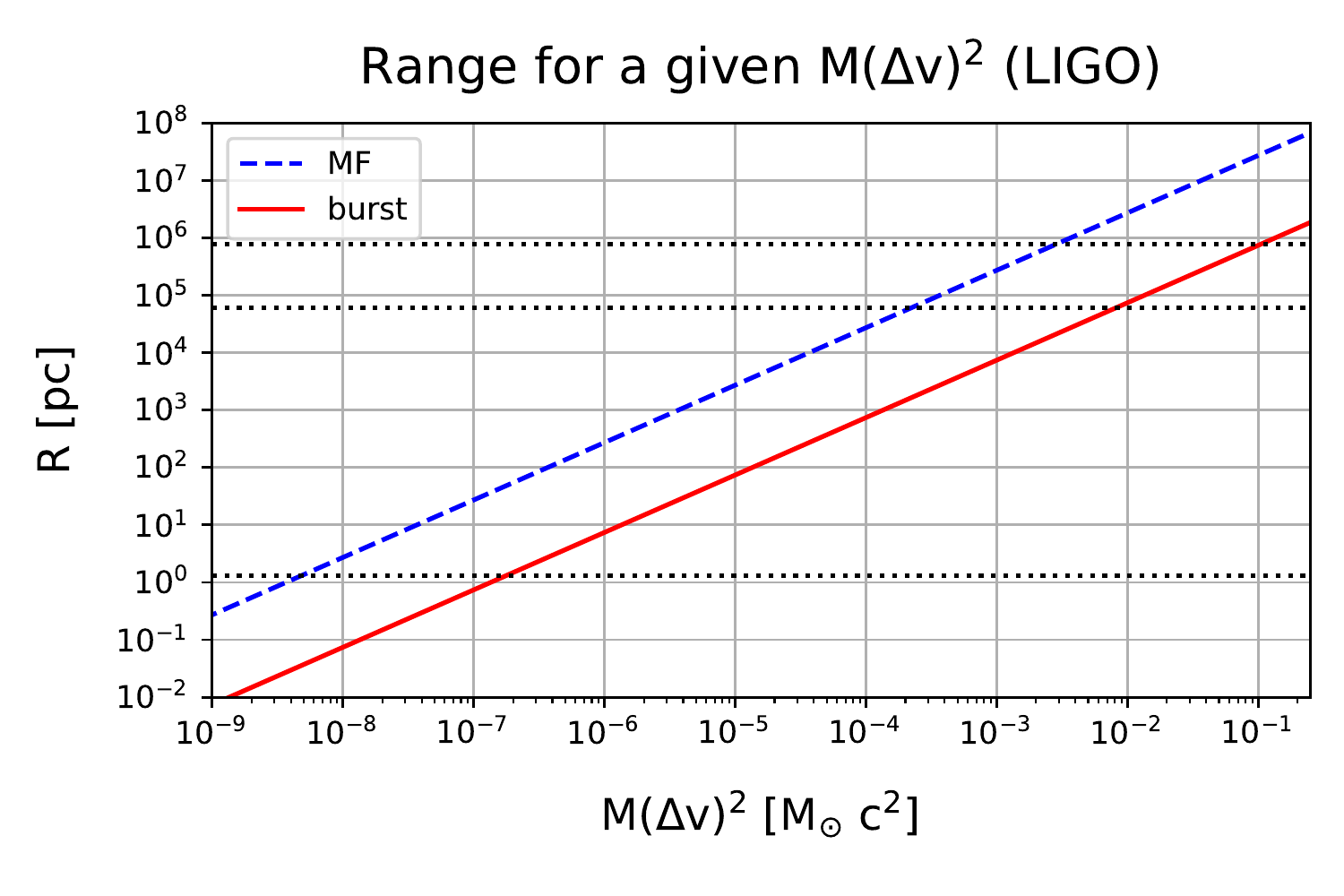}
\includegraphics[width=1.0\linewidth]{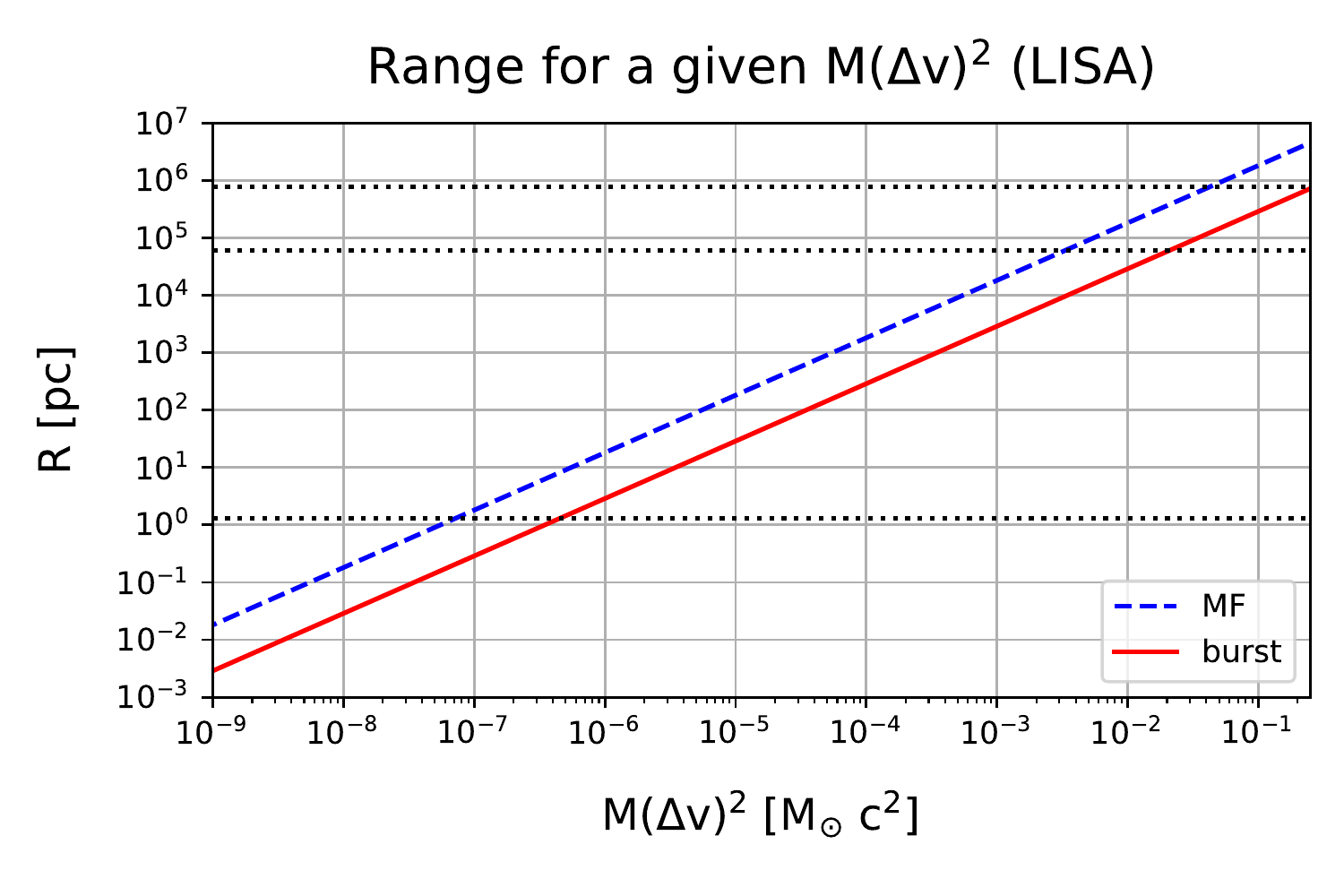}
\caption{Range $R$ for the signal with parameters $M\Delta v^{2} = [10^{-9}\,{\rm M}_{\odot}(c/\text{s})^{2}, 0.25\,{\rm M}_{\odot}c^{2}]$ to be detected by LIGO (Top) or LISA (Bottom) using either a burst or MF search. The regions under the curves yield the parameter sets $\left\{M, \Delta v, R\right\}$ for which the signal would have been detected (LIGO burst) or can be detected (MF, LISA burst). The dotted horizontal lines give, in increasing order, the distance to the nearest star outside of our solar system, the distance to the edge of our Galaxy, and the distance to the Andromeda Galaxy.}
\label{fig:6}
\end{figure}

 
The signal ranges using both burst and MF search methods can be seen in Figure \ref{fig:6} for LIGO and LISA. We look at signals with parameters $M$ and $\Delta v$ ranging from $M\Delta v^{2} = [10^{-9}\; M_{\odot}c^{2}, \;0.25\; M_{\odot}c^{2}]$. The right-hand limit was chosen so that the parameter space ranges up to a solar mass object achieving a velocity that satisfies the mildly relativistic assumption of (\ref{eqn:en_density}). The left-hand bound was chosen such that the range of the maximal curve (LIGO MF search) was approximately $10^{-1}$ pc, around and below which the Newtonian contribution to the strain may begin to dominate. We leave determining exactly where this transition occurs for a future study. Since the x-axis is the change in kinetic energy of the object if the mass $M$ is approximately constant, it is instructive to quantify these bounds in terms of rest-mass conversions of astrophysical objects. The left-hand bound $10^{-9}\,{\rm M}_{\odot} c^{2}$ then corresponds to the rest-mass of an object about one-hundredth the mass of the Moon being converted into kinetic energy, while the right-hand bound corresponds to the rest-mass conversion of a star a quarter of the mass of the Sun. 
\par 
For a given curve, the area underneath gives the parameter space for which detection is possible. For LIGO burst searches in particular, since the SNR is indifferent to the shape of the signal, the area under the burst search curve represents parameter sets $\{M, \Delta v, R\}$ that \textit{would have been detected} by existing burst search pipelines. In other words, these parameter sets have likely not occurred while LIGO was online.
\par 
To summarize the results from Figure~\ref{fig:6}, the largest ranges are given by the LIGO MF search. On the higher end of this plot, we see that \textsc{RAMAcraft} with masses $10^{-3}\,{\rm M}_{\odot}$ (around a Jupiter mass) undergoing a change of velocity of $\Delta v = (0.1)^{1/2} c \sim 0.3\,c$ are detectable up to $10 - 100\,{\rm kpc}$, which covers up to every star in the Milky Way. Therefore, LIGO MF searches can scan our entire Galaxy for Jupiter and super-Jupiter masses achieving similar changes in velocity. On the lower end of the range, we see that Moon-scale masses $\sim 10^{-7}\,{\rm M}_{\odot}$ undergoing the same change in velocity are detectable up to about $1 - 10$\,pc, or around the distance to the nearest stars, such as Proxima Centauri. The results for this change in velocity for these masses and others can be seen in Table \ref{tab:1}\footnote{For the Vesta asteroid, we ignore the Newtonian contribution to the strain and assume that the object does not change its distance to the Earth very much over the data sample length.}. 
\par 
Similarly, using the results from the LIGO burst search, we can rule out the occurrence of certain \textsc{RAMAcraft} (while LIGO was online) by the lack of burst detections. On the upper end of the range, we can rule out 10$^{-1} \rm M_{\odot}$ objects achieving $\Delta v \sim 0.3c$ throughout the entire Galaxy or a Jupiter mass out to about 1\,kpc. On the lower end, we can rule out 10 Moon-mass objects achieving the same speed at around 1\,pc. Since the range is proportional to the mass, we can interpolate between these two regimes by taking $M \to aM$ and $R\to aR$. Similarly, we can interpolate between different $\Delta v$ by taking $\Delta v \to a\Delta v$ and $R \to a^{2}R$. 
\begin{table}
\caption{Ranges for various masses with $\Delta v \sim 0.3 c$. \label{tab:1}}
\centering
  \begin{tabular}{l l l l}
    Object Class & Mass [$\rm M_{\odot}$] & $R_{\rm LIGO, MF}$ [pc] & Within range of:\\
    \hline
    Vesta asteroid   & $10^{-10}$ & $10^{-3} - 10^{-2}$ & Our solar system\\
    \hline
    Moon mass         & $10^{-7}$ & $1 - 10$            & Nearest stars\\
    \hline
    Super-Earth mass & $10^{-5}$ & $10^{2} - 10^{3}$ & Our solar neighborhood\\
    \hline
    Jupiter mass      & $10^{-3}$ & $10^{4} - 10^{5}$ & Nearly every star \\ & & & in the Milky Way\\
    \hline
    Lowest-mass stars    & $10^{-1}$ & $10^{6} - 10^{7}$ & To the Andromeda\\ & & & Galaxy and beyond\\
    \hline
    Solar mass        & $1$       & $10^{7} - 10^{9}$ & Cosmological scales\\
  \end{tabular}
  \label{Tab:Tcr}
\end{table}
\par 
An important distinction remains across different values of $T$ in that the approximation of constant $R$ (\ref{eqn:const_R}) is more easily satisfied for objects with smaller acceleration periods $T$. Since our results assume that $R$ is constant, it is important to discern which parameters are valid for the different regions of the plots. We can do this by plotting the curve
\begin{equation}
R = 5\Delta v T
\end{equation}
for varying values of $T$ and masses $M$ on the $M\Delta v^{2}$ domain in Figure \ref{fig:6}. Points above these curves satisfy the approximation (\ref{eqn:const_R}) to at least a factor of 10. Quoting the numbers here, the results from Figure \ref{fig:6} for the LIGO MF search are accurate for Jupiter-mass objects for the entire plot domain for acceleration periods $T \leq 10^{10}$\,s (over 300 years), while the results are accurate for the region of interest (10 - 100\,kpc) for $T \leq 10^{12}$\,s (over 30,000 years). The results for Moon-mass \textsc{RAMAcraft} are accurate for $T \leq 10^{8}$\,s (over 3 years). For burst searches, the results for Jupiter-mass objects are accurate for the entire plot domain for $T \leq$ 10$^{8}$\,s (over 3 years) and for $T \leq 10^{10}$\,s (over 300 years) for the region of interest ($\sim$ 1\,kpc). The results for ten Moon-mass $10^{-6}\,{\rm M}_{\odot}$ objects are valid for $T \leq 5\times 10^{6}$\,s (about 2 months). To summarize, the results for the LIGO MF search are quite robust to the assumption (\ref{eqn:const_R}), though loosening this requirement may be warranted for burst search results. We leave this for future studies.
\subsubsection{Comparing LIGO and LISA}\hfill \\
\par
From Figure \ref{fig:6}, we can see that LIGO and LISA perform comparably for both burst and MF searches, but that LIGO edges LISA in both cases, namely by a factor of a few for burst searches and around ten for MF searches. 
\par 
This may be counter-intuitive because, for example, the signal spectrum scales like $f^{-1}$, and LISA is more sensitive for lower frequencies. We can see why this is the case from a simple analytical argument. We aim to compute a ratio of ranges for LIGO and LISA:
\begin{equation}
\label{eqn:rdet}
    r_{\rm det} = \frac{R_{\text{LIGO}}}{R_{\text{LISA}}}
\end{equation}
Then for both detectors, the SNR squared accumulated at one point is proportional to 
\begin{equation}
\label{eqn:d_burst}
    \Delta
    \left(\frac{\rho^{2}_{\rm b}}{\sigma^{2}_{\rm b}}\right)
    \propto
    \frac{|h(f)|^{2}}{|n(f)|^{2}}
\end{equation}
in the case of burst searches (see (\ref{eqn:pb}-\ref{eqn:noise_burst_SNR})) and 
\begin{equation}
\label{eqn:d_MF}
    \Delta
    \left(\frac{\rho^{2}_{\rm MF}}{\sigma^{2}_{\rm MF}}\right)
    \propto
    \frac{|h(f)|^{2}}{n(f)\tilde{h}(f)}
    \sim
    \frac{h(f)}{n(f)}
\end{equation}
in the case of MF searches (see (\ref{eqn:pmf}-\ref{eqn:sigmf})), where the last relation holds up to a tripling of the complex phase of the strain spectrum:
\begin{equation}
z\cdot
\frac{z}{\tilde{z}}
=
re^{i\phi}\cdot
\frac{re^{i\phi}}{re^{-i\phi}} 
=
re^{3i\phi}
\end{equation}
Then if we extract the $R$ dependence of each spectrum by $R\cdot h(f) = h_{0}(f)$, we find that each statistic scales like
\begin{equation}
\begin{split}
 &\Delta
    \left(\frac{\rho^{2}_{\rm b}}{\sigma^{2}_{\rm b}}\right)
    \propto
    \frac{1}{R^{2}}
    \frac{|h_{0}(f)|^{2}}{|n(f)|^{2}}\\
&\Delta
    \left(\frac{\rho^{2}_{\rm MF}}{\sigma^{2}_{\rm MF}}\right)
    \propto
    \frac{|h(f)|^{2}}{n(f)\tilde{h}(f)}
    \sim
    \frac{h(f)}{n(f)}
    =
    \frac{1}{R}
    \frac{h_{0}(f)}{n(f)}
\end{split}
\end{equation}
and we again find that the range $R$ scales like $R \propto (\rho_{\rm b} / \sigma_{\rm b})$ or $R \propto (\rho^{2}_{\rm MF} / \sigma^{2}_{\rm MF})$, as we saw in the methodology section.
Then moving from LIGO to LISA, our overall sensitivity drops from about $5\times10^{-24}$\,Hz$^{-1/2}$ to $10^{-20}$\,Hz$^{-1/2}$, and the sensitivity band minimum shifts from around $100$\,Hz to $5 \times 10^{-3}$\,Hz. Since the signal spectrum scales like $f^{-1}$, the signal spectrum will be larger in LISA's sensitivity band than LIGO's. Since these effects will apply for each SNR contribution (\ref{eqn:d_burst}) and (\ref{eqn:d_MF}), we can extrapolate this effect to the total SNR as well, yielding the following proportionality for $r_{\rm det}$:
\begin{equation}
\label{eqn:r0}
    r_{\rm det} \propto
    \left(\frac{5\times 10^{-3}\,\text{Hz}}{100 \,\text{Hz}}\right)
    \left(\frac{10^{-20}\,\text{Hz}^{-1/2}}{5\times 10^{-24}\,\text{Hz}^{-1/2}}\right)
\end{equation}
for both search methods. The proportionality (\ref{eqn:r0}) is the same for both search methods due to the fact that $R \propto \rho_{\rm b} / \sigma_{\rm b}$ and $R \propto \rho^{2}_{\rm MF} / \sigma^{2}_{\rm MF}$, respectively. However, there are two more effects that we need to take into account. The first is for burst searches, namely that since the burst SNR scales like $\tau^{-1/2}$, LIGO will receive a gain in range by virtue of the fact that it can access its sensitivity band with a smaller $\tau$. We then find that the ratio (\ref{eqn:rdet}) for burst searches is
\begin{equation}
\label{eqn:r_det_b}
    r^{(b)}_{\rm det}
    \approx 
    \left(\frac{5\times 10^{-3}\,\text{Hz}}{100\,\text{Hz}}\right)
    \left(\frac{10^{-20}\,\text{Hz}^{-1/2}}{5\times 10^{-24}\,\text{Hz}^{-1/2}}\right)
    \left(
    \frac{\tau_{\text{LISA}}}{\tau_{\text{LIGO}}}
    \right)^{1/2}
\end{equation}
Then from our results for the burst search, we chose $\tau_{\text{LISA}} = 200$\,s and $\tau_{\text{LIGO}} = 0.1$\,s. Plugging these values in then predicts $r^{(b)}_{\rm det} \approx 4$. Using the data from Figure~\ref{fig:6}, we find that $r^{(b)}_{\rm det} \approx 3$, so our results are in good agreement with this prediction. 
\par 
For MF searches, recall from (\ref{eqn:MF_SNR_tau}) that the noise correlation $\sigma_{\rm MF}$ receives an additional scaling of $\tau^{-1/4}$, which we conjecture is due to the changing number of points in the noise correlation computation. Then for a given frequency spacing $\tau^{-1}$, since the LIGO sensitivity band sits at around 100\,Hz vs 5\,mHz for the case of LISA, the LIGO noise correlation computation contains approximately 100 / 0.005 = 20,000 times as many points as LISA. One may try to overcome this effect by increasing the sample length for LISA, but this will not increase the SNR by (\ref{eqn:MF_SNR_tau}). Then since the range scales like $R \propto \rho^{2}_{\rm MF} / \sigma^{2}_{\rm MF}$, this effect enters the ratio expression like 
\begin{equation}
\begin{split}
\label{eqn:r_det_MF}
    r^{({\rm MF})}_{\rm det}
    &\approx 
    \left(\frac{5\times 10^{-3}\,\text{Hz}}{100 \,\text{Hz}}\right)^{1/2}
    \left(\frac{10^{-20}\,\text{Hz}^{-1/2}}{5\times 10^{-24}\,\text{Hz}^{-1/2}}\right)
    \approx 
    14
\end{split}
\end{equation}
which again matches very closely what we get using the data from Figure~\ref{fig:6}, namely $r^{({\rm MF})}_{\rm det} \approx 15$.
\subsubsection{Burst vs MF Search}\hfill \\
\par 
In Figure \ref{fig:6}, we can see explicitly that the MF search method is more sensitive than the band-pass burst search, as expected. From the results just discussed, we expect the benefit of moving from a burst to a MF search for LIGO and LISA to defer by the ratio $r^{(\rm MF)}_{\rm det} / r^{(\rm b)}_{\rm det} \approx 3$.
Empirically, we find that the ratio of ranges using a MF vs burst search is about 36 for the case of LIGO and about 6 for the case of LISA, yielding a relative factor of about 6 for the gain in switching from a burst to a MF search for LIGO compared to LISA, which agrees reasonably with our prediction. The discrepancy from prediction may be explained by the variations in $\rho_{\rm MF} / \sigma_{\rm MF}$ for different values of $\tau$ detailed in Appendix \ref{app:d}.
\section{Discussion}
\label{sec:Discussion}
In this section, we discuss how sensitivity improvements for low frequencies in particular will benefit the search for \textsc{RAMAcraft}, ways in which our analysis can be generalized to other acceleration mechanisms, and potential natural sources that can mimic a \textsc{RAMAcraft} signal.
\subsection{Sensitivity Improvements: Comparing Other Detectors}
The results (\ref{eqn:r_det_b}) and (\ref{eqn:r_det_MF}) are not to say that low-frequency detectors aren't more useful for detecting \textsc{RAMAcraft}. For the case of LISA, the deficit in overall sensitivity as compared to LIGO is too large to take advantage of the low-frequency sensitivity. There are, however, low-frequency detectors for which this is not the case, such as DECIGO and the Big Bang Observer (BBO). For the case of DECIGO, the sensitivity band minimum sits at around 0.5 Hz with a sensitivity of about $10^{-24}$\,Hz$^{-1/2}$ \citep{Moore_2015}. Assuming that (\ref{eqn:r_det_b}) and (\ref{eqn:r_det_MF}) continue to hold, plugging these numbers in and using $\tau_{\text{DEC}} = 20$\,s yields both $r^{(b)}_{\rm det}$ and $r^{({\rm MF})}_{\rm det}$ $\approx 0.01$. DECIGO searches will therefore be 100 times more sensitive than LIGO, increasing the search volume by a factor of 10$^{6}$. Similarly, the sensitivity band of the BBO sits at around 0.1 Hz, also with a sensitivity of about $10^{-24}$\,Hz$^{-1/2}$ \citep{Moore_2015}. Plugging these values in with $\tau_{\text{BBO}} = 40$\,s yields  $r^{(b)}_{\rm det}$ and $r^{({\rm MF})}_{\rm det}$ $\approx 0.005$, making the range of the BBO even better than DECIGO by about a factor of 2.
\par 
Another promising detection method to consider is the use of Pulsar Time Arrays (PTAs), since their sensitivity bands fall within the nHz regime \citep{PTAs, realistic_curves_PTA}. Although the sensitivity curves presented in \citet{realistic_curves_PTA}  may look reasonably sensitive, these plots are of the \textit{characteristic strain}, which carries a factor of $f^{1/2}$ compared to $S_{n}(f)^{1/2}$. So even though the sensitivity band minimum occurs around 5\,nHz, the overall sensitivity of about $10^{-10}$ Hz$^{-1/2}$ is too low to provide a benefit as compared to LIGO. Plugging these values into (\ref{eqn:r_det_b}) and (\ref{eqn:r_det_MF}) and using $\tau_{\text{PTA}} = 10^{9}$\,s (around 30 years), we find that $r^{(b)}_{\rm det}$ and $r^{({\rm MF})}_{\rm det}$ $\approx 10^{8}$. However, because the sensitivity band is on the order of nHz, PTAs can be less sensitive than LIGO by a factor of about $10^{5}$ and still provide a benefit $r_{\rm det} < 1$. Therefore,  sensitivity improvements to PTAs could significantly improve the feasibility of detecting \textsc{RAMAcraft}. 
\subsection{Sensitivity Improvements: Longer Acceleration Periods}
The preceding analysis is predicated on the assumption that the entire acceleration period is contained within the data sample, or in other words, that the `perceived change in velocity' does not change with the data sample length $\tau$. However, this very well may not be the case. More to the point, detectors that can record data for longer periods of time without interruption may be much more sensitive to \textsc{RAMAcraft} with longer acceleration periods, as is evident from the scaling laws (\ref{SNRb_tau_bigT} - \ref{eqn:MF_SNR_tau_bigT}). There is good reason to believe that LISA and other space-based detectors will be capable of longer uninterrupted data samples than LIGO due to the fact that LISA will be isolated in space, whereas LIGO experiences the perturbing influence of ground-based noise. A reasonable prediction would be that LISA will be capable of uninterrupted samples of up to 1 week, whereas LIGO's capacity will be approximately 1 day (Professor Neil Cornish, private communication). 
\par 
The impact that this effect would have on $r_{\rm det}$ is clear from (\ref{SNRb_tau_bigT} - \ref{eqn:MF_SNR_tau_bigT}). For burst searches, the scaling (\ref{SNRb_tau_bigT}) will overpower the third factor in (\ref{eqn:r_det_b}) and yield
\begin{equation}
\label{eqn:rdet_b_bigT}
\begin{split}
    r^{(b)}_{\rm det}(T > t_{2})
    \approx 
    &\left(\frac{5\times 10^{-3} \text{Hz}}{100 \; \text{Hz}}\right)
    \left(\frac{10^{-20} \; \text{Hz}^{-1/2}}{5\times 10^{-24} \; \text{Hz}^{-1/2}}\right)\\
    &\times \left(
    \frac{\tau_{\text{LISA}}}{\tau_{\text{LIGO}}}
    \right)^{-1}
\end{split}
\end{equation}
which gives $\approx 0.014$ for $\tau_{\text{LISA}} = 1$\,week and $\tau_{\text{LIGO}} = 1$\,day. For a MF search, keeping in mind that $R \propto \rho^{2}_{\rm MF} / \sigma^{2}_{\rm MF}$, we should get
\begin{equation}
\label{rdet_mf_bigT}
\begin{split}
    r^{(\text{\rm MF})}_{\rm det}(T > t_{2})
    \approx 
    &\left(\frac{5\times 10^{-3} \text{Hz}}{100  \text{Hz}}\right)^{\frac{1}{2}}
    \left(\frac{10^{-20} \text{Hz}^{-1/2}}{5\times 10^{-24}  \text{Hz}^{-1/2}}\right)\\
    &\times 
    \left(
    \frac{\tau_{\text{LISA}}}{\tau_{\text{LIGO}}}
    \right)^{-2}
\end{split}
\end{equation}
which gives $\approx 0.29$. Therefore, LISA will be superior to LIGO for both burst and MF searches for acceleration periods exceeding 1 week. It seems reasonable to expect that a large proportion of \textsc{RAMAcraft} trajectories will accelerate for at least 1 week, in which case LISA will be superior to LIGO at searching for these trajectories. 
\par 
Similarly, if DECIGO is also capable of 1-week-long data samples, it would provide a benefit of $r^{(\text{\rm b})}_{\rm det}(T > t_{2}) \approx 10^{-4}$ and $r^{(\text{\rm MF})}_{\rm det}(T > t_{2}) \approx 3\times 10^{-4}$, which would increase the search volume of MF and burst searches by 10$^{11}$ and 10$^{12}$, respectively. Similarly, the BBO would provide a benefit of  $r^{(\text{\rm b})}_{\rm det}(T > t_{2}) \approx 3\times 10^{-5}$ and $r^{(\text{\rm MF})}_{\rm det}(T > t_{2}) \approx 10^{-4}$, which would increase the search volume of MF and burst searches by 10$^{12}$ and 10$^{14}$, respectively.
\par 
Finally, the most extreme examples of uninterrupted data collection come from PTAs, which can record for up to several years without interruption. Plugging in the PTA parameters and $\tau_{\text{PTA}} = 10$\,years into (\ref{eqn:rdet_b_bigT}) and (\ref{rdet_mf_bigT}), we find that $r^{(\text{b})}_{\rm det}(T > t_{2}) \approx$ 0.27 and $r^{(\text{\rm MF})}_{\rm det}(T > t_{2}) \approx$ 10.62. So, PTAs would be more sensitive to bursts from the get-go and could be about $10^{12}$ times less sensitive than LIGO and still provide a benefit for MF searches. Therefore, future sensitivity improvements to PTAs may be of particular value for detecting \textsc{RAMAcraft} with acceleration periods spanning a year and beyond.
\begin{figure}
\centering
\includegraphics[width=1.0\linewidth]{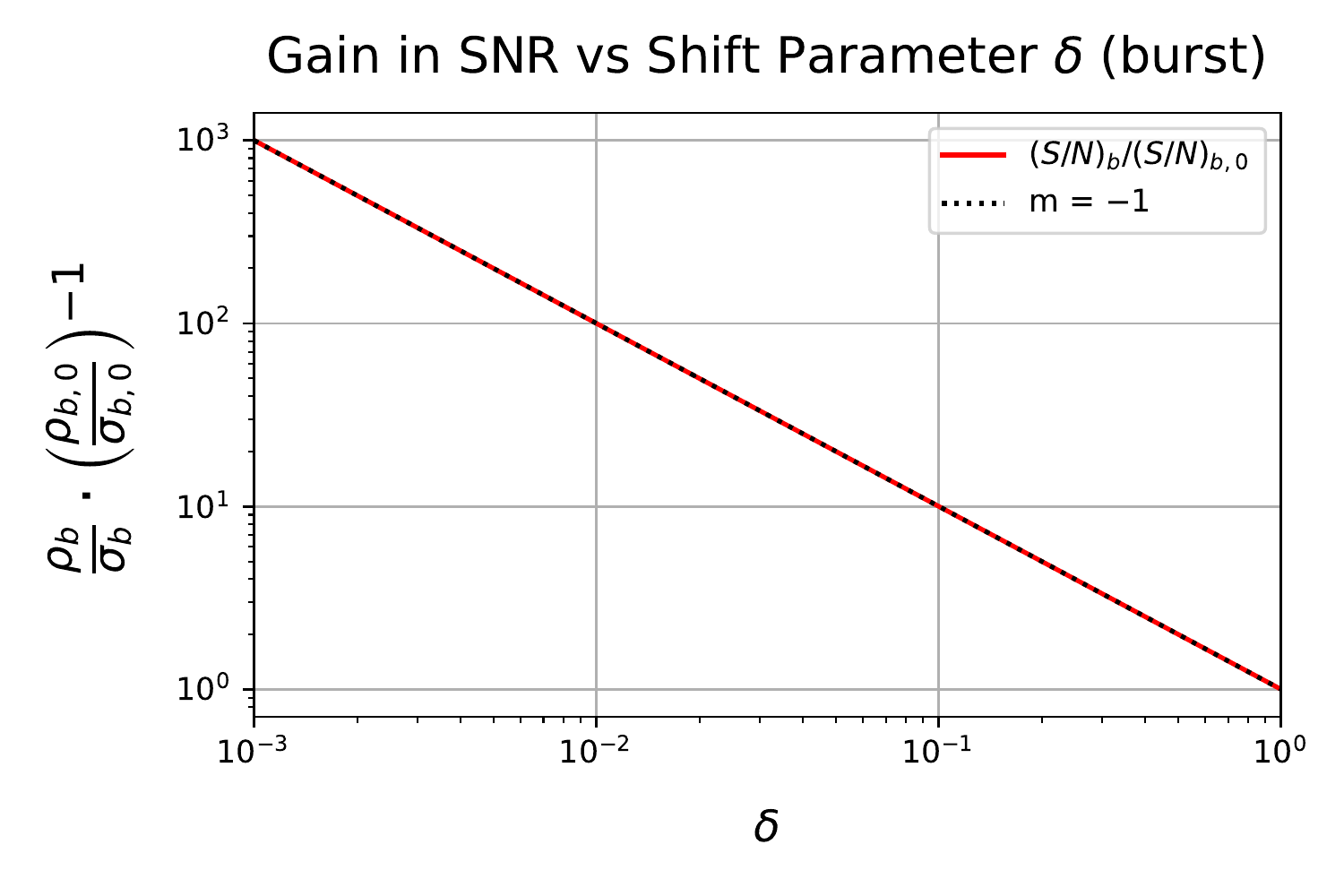}
\includegraphics[width=1.0\linewidth]{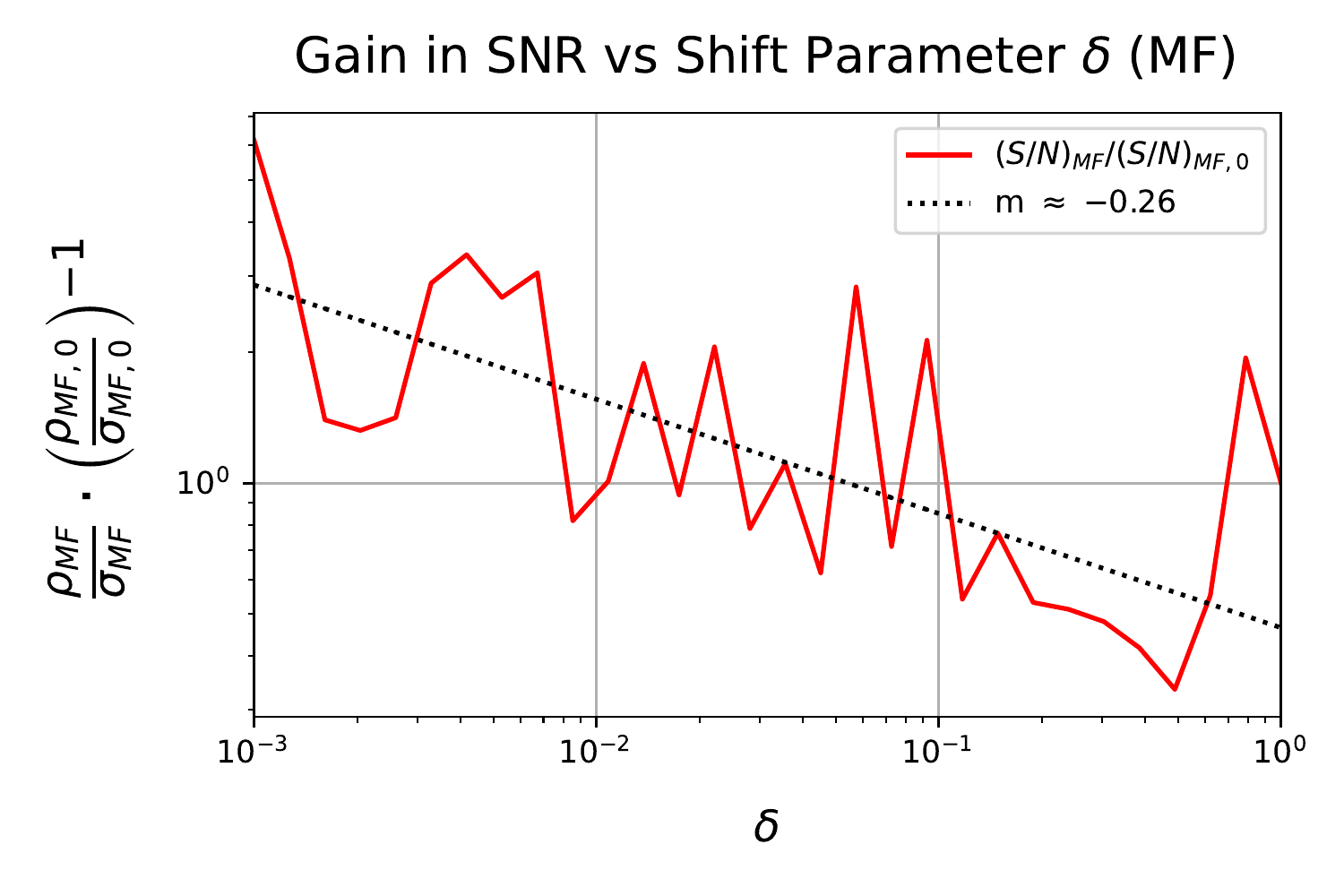}
\caption{Sensitivity gain achieved by shifting over the LISA sensitivity curve by a factor $\delta < 1$. The quantity $\rho/\sigma\cdot (\rho_{0}/\sigma_{0})^{-1}$ is the ratio of the SNR computed when shifting the sensitivity curve by the factor $\delta$ to the SNR computed when the curve is not shifted at all. The scaling behavior $\delta^{-1}$ for $\rho_{\rm b} / \sigma_{\rm b}$ matches nearly perfectly, while the scaling behavior $\delta^{-0.25}$ for $\rho_{\rm MF} / \sigma_{\rm MF}$ has some oscillations about the behavior that we expect, likely due to inconsistencies associated with computing $\sigma_{\rm MF}$.}
\label{fig:7}
\end{figure}
\subsection{Sensitivity Improvements: Overall Sensitivity and Shifting the Sensitivity Band}
\par
So far, we have examined sensitivity improvements that come from moving between different detectors. Here, we assess the benefit of sensitivity improvements for a given detector, namely improvements to the overall sensitivity and shifting the sensitivity band. 
\par 
We can quantify improvements to the overall sensitivity of a given detector by simply scaling its sensitivity curve: 
\begin{equation}
S_{n}(f) \to \alpha S_{n}(f)
\end{equation} for $\alpha < 1$. Given  (\ref{eqn:pb} - \ref{eqn:sigmf}), this would result in an increase of the burst and MF SNRs by a factor of $\alpha^{-1}$ and $\alpha^{-1/2}$, respectively, both of which would increase the range by a factor of $\alpha^{-1}$.
\par  
Similarly, we can quantify a sensitivity improvement to lower frequencies by shifting the sensitivity curve like
\begin{equation}
  S_{n}(f) \to S_{n}(\delta f)  
\end{equation} for $\delta < 1$. The effect that this would have on the SNR is also straightforward to work out. For a given signal $h(f)$, the effect of shifting the sensitivity band of $S_{n}$  by a factor $\delta$ is to accumulate the majority of the SNR for frequencies where $h(f)$ is larger by a factor of $\delta^{-1}$. This will result in $\rho_{\rm b} / \sigma_{\rm b}$ increasing by a factor of $\delta^{-1}$. For MF searches,
$\rho_{\rm MF} / \sigma_{\rm MF}$ will increase by $\delta^{-1/2}$ due to the increase in $h(f)$. However, $\rho_{\rm MF} / \sigma_{\rm MF}$ will also decrease by a factor $\delta^{1/4}$ due to a decrease in the number of points used to calculate the noise correlation $\sigma_{\rm MF}$ (see Appendix \ref{app:d}). Thus, the total increase in the MF SNR $\rho_{\rm MF} / \sigma_{\rm MF}$ should be by a factor of $\delta^{-1/4}$. Therefore, the range will increase by a factor of $\delta^{-1}$ for burst searches and a factor of $\delta^{-1/2}$ for MF searches.  
This effect can be seen in Figure~\ref{fig:7}.
\subsection{Sensitivity Improvements: Elongating Data Samples Past the Acceleration Period}
As we mentioned in Section \ref{ref:Sec2LinAcc}, assuming that the approximation (\ref{eqn:const_R}) is valid for $\tau > T$, we can extend the $f^{-1}$ scaling of the spectrum past $T^{-1}$ by increasing the data sample length $\tau$. This may be useful to take advantage of the low-frequency sensitivity of some hypothetical detector. Assuming that the acceleration duration is entirely contained in the data sample, elongating $\tau$ will lengthen the portions of constant velocity contributions. Since the integration bounds enter these contributions only in their phase argument, this does not result in any scaling behavior. However, as we increase $\tau$, we do gain access to lower frequencies in our spectrum $f = k\tau^{-1}$.
\par 
Since our spectrum scales like $f^{-1}$, increasing data sample times may boost the range for these objects for some combination of data sample length and detector sensitivity band. The extent to which we can extend $\tau$ over the constant velocity contributions will depend on the duration said contributions remain in the detector’s field of view. More to the point, since this paper assumes that the object is approximately a constant distance from the detector, we can only extend $\tau$ for a duration less than or equal to the point where this approximation is no longer valid. In order to determine this duration, we can define the `time in view' $T_{R}$ for the object at a given distance from the detector:
\begin{equation}
    T_{R} = \frac{R}{10\langle v\rangle}
\end{equation}
where we claim that the constant distance approximation is valid for objects that travel 10\% of their initial distance from the detector.
Then $\langle v\rangle$ is the average velocity of the signal over the data sample and is given by
\begin{equation}
\begin{split}
    \langle v \rangle 
    &=
    \frac{1}{\tau}
    \int_{t_{1}}^{t_{2}}v(t) dt\\
    &=
    v_{0} + \frac{t_{2} - T/2}{\tau}\Delta v
\end{split}
\end{equation}
where again $\Delta v = AT$.
The result clearly depends on which velocities are captured by the data sample, though we can gauge the extent to which we can extend our data sample for objects that undergo a very large change in velocity compared to their initial velocity $\Delta v \gg v_{0}$. In this case, assuming that a reasonable portion of the final velocity is present in the data sample, the average velocity is $\langle v\rangle \approx \Delta v$, so $T_{R}$ is then given by
\begin{equation}
    T_{R}
    \approx
    \frac{R}{10\Delta v}
\end{equation}
At the lower end of the ranges in Figure~\ref{fig:6}, a \textsc{RAMAcraft} located 1\,pc away undergoing a change in velocity of $\Delta v \sim 0.3\,c$ has a time in view of about 10.9 years. At the upper end of the range, a \textsc{RAMAcraft} located 100 \,kpc away achieving the same change in velocity has a time in view of about 10$^{6}$ years. So towards the upper end of the ranges in Figure~\ref{fig:6}, we can extend $\tau$ practically indefinitely to try to take advantage of some hypothetical low-frequency detector. Towards the lower end of the range, we can still extend $\tau$ quite considerably, which would almost certainly cover the capabilities of ground-based and space-based interferometers, as well as a substantial portion if not the entirety of the detection lengths of PTAs. 
\par 
Of course, it is not prudent to extend the data sample to frequencies $\tau^{-1}$ to the left of the detector ASD sensitivity band (unless a detector is built whose left-hand sensitivity band boundary scales weaker than $f^{-1}$). The benefit just described is therefore contingent on the improvement of detector sensitivities to lower frequencies. 
\par 
Finally, we note that this is indeed a physical effect and not an artifact of data sample parameters. The boost in range by extending the $f^{-1}$ behavior of the spectrum into some low-frequency sensitivity band is afforded by the fact that the \textsc{RAMAcraft} and the gravitational memory contributed by its change in velocity is in view of the detector for the requisite amount of time. 
\subsection{Including More General Objects}
In our analysis, we have assumed that the \textsc{RAMAcraft} move in a straight line by some propulsive mechanism, and our frequency analysis furthermore assumes that their accelerations are constant. Objects that would circumvent these specifications include:
\begin{enumerate}

    \item 
    \label{gen0}
    Objects with a non-linear trajectory
    
    \item 
    \label{gen1}
    Objects with a non-constant acceleration
    
    \item 
    \label{gen3}
    Hypothetical objects that accelerate by some non-propulsive mechanism
\end{enumerate}
To include these generalizations in a search network, we need only compute their GW signal shapes and include them in the catalog of MF templates, which we leave for future studies.
\par 
In terms of the first two generalizations, we can at least consider some regime in which the acceleration and trajectory of the object are approximately constant and straight, respectively. For example, highly eccentric BBH mergers, which may evolve dynamically within dense star clusters \citep{eccentric_mergers, EccBH}, can effectively be seen as a mass falling straight into a gravitational well for a portion of the orbit for eccentricities close to 1. Specifically, for the case where some mass $m$ has  an initial separation $R$ from a much larger mass $M$, we have that
\begin{equation}
    \frac{GMm}{R} = \frac{1}{2}mv^{2}
\end{equation}
assuming relatively small velocities prior to the formation of the orbit. Then dividing both sides by $mc^{2}$, we find that
\begin{equation}
\label{eqn:R_schwarz}
    \frac{R_{\rm s}}{R}
    = 
    \left(
    \frac{v^{2}}{c^{2}}
    \right)
    =
    \beta^{2}
\end{equation}
or in other words, objects falling into a gravitational well achieve speeds comparable to the speed of light only near the Schwarzschild radius $R_{\rm s} = 2GM/c^{2}$ of other massive objects. \begin{figure}
    \centering
    \includegraphics[width = 1.0\linewidth]{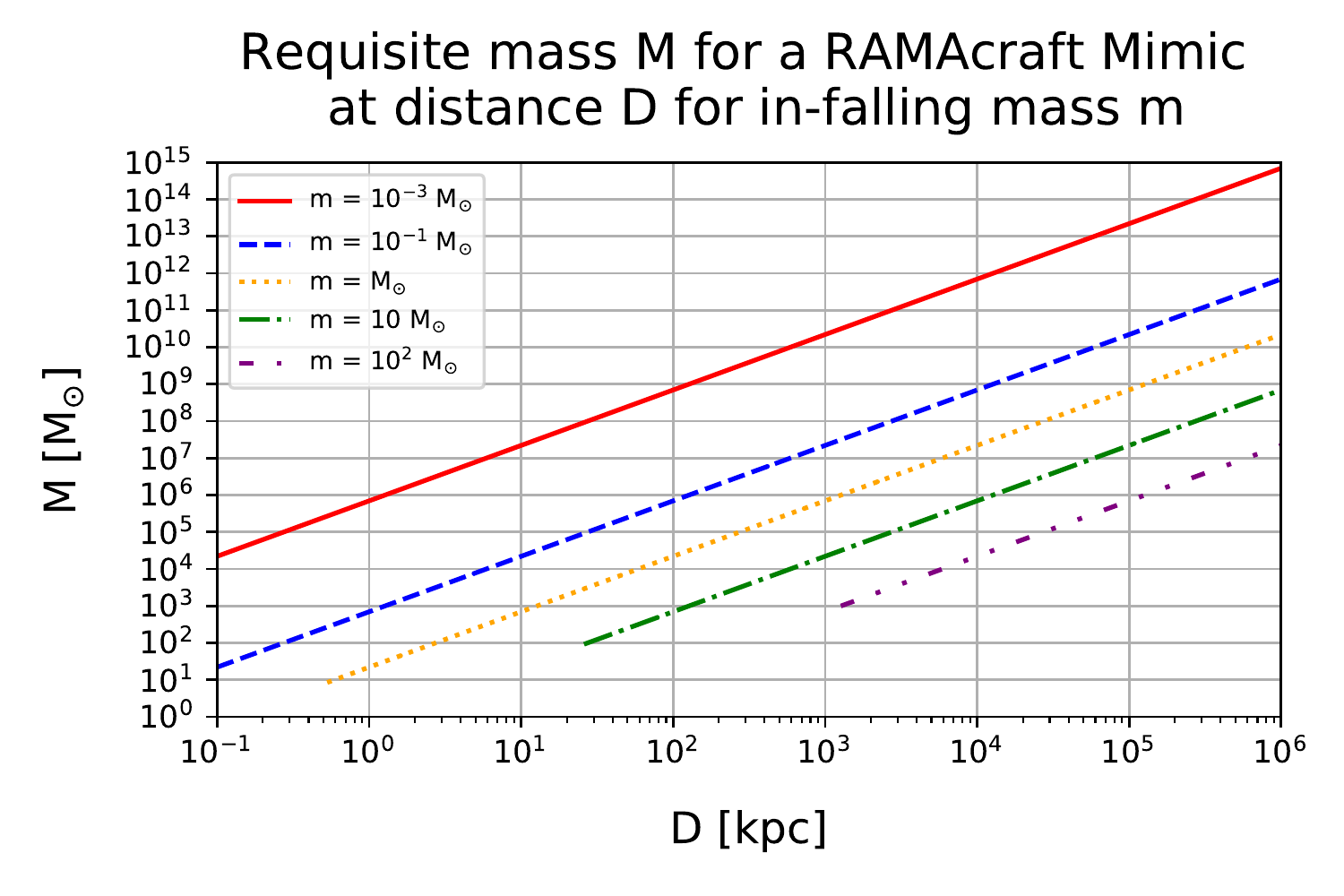}
    \includegraphics[width = 1.0\linewidth]{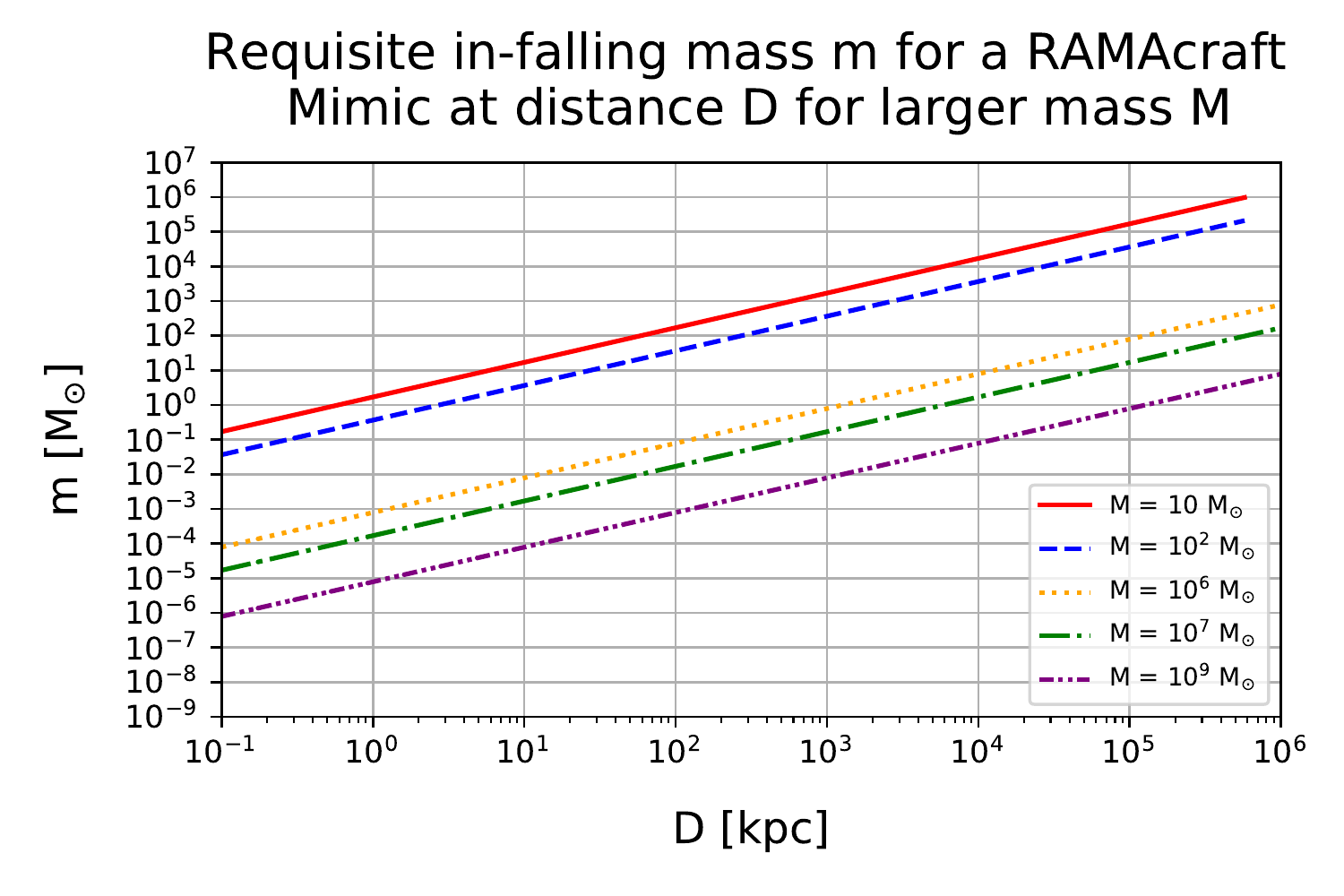}
    \caption{Masses required for a detectable highly-eccentric merger capable of mimicking a \textsc{RAMAcraft} signal over a 1$\,$h data sample for a given distance to the merger $D$ and for a given in-falling mass $m$ (Top) or larger mass $M$ (Bottom). Some of the plots are cut off where the in-falling mass exceeds one-tenth of the larger mass $M$, in which case the approximation that $M$ is not affected by the smaller mass is no longer valid. We omit intermediate-mass BHs $10^{2}-10^{5}\,\rm M_{\odot}$ due to lack of prevalence.}
    \label{fig:8}
\end{figure}
\par
Now we will examine cases where these highly-eccentric orbits can mimic the signals that we examine in Section \ref{sec:Results}. In particular, we want to examine cases where the acceleration of the smaller mass $m$ is approximately constant. For $M\gg m$, this will be the case for the time period:
\begin{equation}
\label{eqn:const_A_infall}
    T^{2}
    \sim
    \frac{R^{3}}{10GM}
\end{equation}
where we are using the condition that the mass $m$ travels 5\% of the separation $R$. 
After this time period elapses, the two masses will either collide with each other or enter a regime in which the separation and the acceleration is no longer constant. In either case, the signal will evolve into something different than the signals that we study in Section \ref{sec:Results}, and the two can be differentiated. Therefore, in order for these highly-eccentric mergers to mimic a \textsc{RAMAcraft} signal with constant acceleration, the time period (\ref{eqn:const_A_infall}) must exceed the length of uninterrupted LIGO data samples. While these sample lengths will vary due to random data glitches, we can get an estimate as to what merger parameters are detectable by choosing a conservative time period $T$, which we do towards the end of this analysis. 
\par 
Then ignoring frequency dependence, we know that the strain from the in-falling object will look like 
\begin{equation}
\label{eqn:hm}
    h_{m}
    \sim
    \frac{2Gm\Delta v^{2}}{Dc^{4}}
    =
    \frac{r_{\rm s}\beta^{2}}{D}
\end{equation}
where $D$ is the distance to the orbit and $r_{s} = 2Gm/c^{2}$ is the Schwarzschild radius of the smaller mass $m$. Then combining (\ref{eqn:R_schwarz}) and (\ref{eqn:const_A_infall}), we have
\begin{equation}
\label{eqn:T}
T \sim
\frac{R_{s}}{c\beta^{3}\sqrt{5}}
\end{equation}
which we can use to solve for $\beta$ and write
\begin{equation}
\label{eqn:hm_final}
    h_{m}
    \sim
    \left(
    \frac{r_{\rm s}}{D}
    \right)
    \left(
    \frac{R_{\rm s}}{cT\sqrt{5}}
    \right)^{2/3}
\end{equation}
We can then compare the strain (\ref{eqn:hm_final}) to those generated by objects on the threshold of detection in Figure \ref{fig:6} and find for what parameters a highly-eccentric merger will be detectable by a LIGO MF search. For example, we can choose a Jupiter-mass \textsc{RAMAcraft} at a conservative distance of 10$\,\rm kpc$ with a change in speed of $\beta \sim 0.3$. With these parameters, the \textsc{RAMAcraft} strain will look like
\begin{equation}
\label{eqn:hJ}
    h_{\rm J}
    \sim 
    \frac{R_{\rm J}}{10\,\rm kpc}
    (0.3)^{2}
\end{equation}
where $R_{\rm J}$ is the Schwarzschild radius of a Jupiter mass. Then taking the ratio of (\ref{eqn:hm_final}) and (\ref{eqn:hJ}), we have that
\begin{equation}
    \frac{h_{m}}{h_{\rm J}}
    =
    \left(
    \frac{m}{10^{-3}\,\rm M_{\odot}}
    \right)
    \left(
    \frac{10\,\rm kpc}{D}
    \right)
    \left(
    \frac{R_{\rm s}}{(0.3)^{3}cT\sqrt{5}}
    \right)^{2/3}
\end{equation}
In order for the merger to be detectable, we need this ratio to be 1 (at the present moment). In this case, we can solve for the larger mass $M$ in terms of $D$ for a given in-falling mass $m$. Doing so yields
\begin{equation}
    M
    =
    \left(
    \frac{D}{10\,\rm kpc}
    \right)^{3/2}
    \left(
    \frac{m}{10^{-3}\,\rm M_{\odot}}
    \right)^{-3/2}
    \left(
    \frac{(0.3c)^{3}T\sqrt{5}}{2G}
    \right)
\end{equation}
Then we can simplify by choosing a conservative sample length for LIGO of $T = 1\,$h:
\begin{equation}
    M
    \approx 
    2.21\times 10^{7}\,{\rm M}_\odot\left(
    \frac{D}{10\,\rm kpc}
    \right)^{3/2}
    \left(
    \frac{m}{10^{-3}\,\rm M_{\odot}}
    \right)^{-3/2}
\end{equation}
The mimic candidates $M$ can be seen in Figure \ref{fig:8} for a given distance to the orbit $D$ and in-falling mass $m$. Likewise, it is also useful to choose the larger mass candidate $M$ and examine what the requisite in-falling mass $m$ would need to be for a detectable signal. In this case, we have
\begin{equation}
    m
    \approx
    7.87\cdot10^{-3}\,{\rm M}_\odot\left(
    \frac{D}{10\,\rm kpc}
    \right) \left(
    \frac{M}{10^{6}\,\rm M_{\odot}}
    \right)^{-2/3}
\end{equation}
which can be seen in Figure \ref{fig:8} as well. Then we can see that if the larger mass $M$ is provided by the plentiful supply of stellar mass BHs $10-100\,\rm M_{\odot}$ in our Galaxy located between $0.1-1\,\rm kpc$, the in-falling mass $m$ will need to be around $10^{-1} - 1\,\rm M_{\odot}$\footnote{Assuming that tidal forces do not disrupt the in-falling mass}. Similarly, if the $M$ candidates are supermassive BHs $10^{6}-10^{9}\,\rm M_{\odot}$ that are prevalent between $10-100\,$million light-years ($3-30\,\rm Mpc$), then the in-falling mass can be smaller at around $10^{-3}-1\,\rm M_{\odot}$. 
\par 
The prevalence of these mergers may be estimated based on the projected rates of extreme-mass-ratio BBH mergers. For example, LISA is expected to observe $10$--$2000$ extreme-mass-ratio mergers per year \citep{EMRIs}. While the majority of such events are expected to have non-extreme eccentricities, it is entirely possible that some orbits may be nearly radial. Therefore, it is important to follow up on potential detections for subsequent increases in acceleration. Furthermore, detectors that are capable of longer uninterrupted data samples will decrease the likelihood that a \textsc{RAMAcraft} signal is mimicked by one of these mergers. 
\par 
In the event that a detected signal is not mimicked by some highly-eccentric orbit or otherwise, the signal may be generated by some mode of transportation satisfying generalization \ref{gen3}. These signals may closely match the strain (\ref{eqn:time_strain}) or have some other shape, depending on the transportation mechanism. Examples proposed in the literature include Warp Drive spacetimes, e.g. \citet{Alcubierre1994}. While energy fluxes can be seen close to the Alcubierre Drive \citep{alcubierre_energy_flux}, none of the warp drive spacetimes proposed thus far emit GWs far from the source. In other words, none of the proposed metrics have an asymptotic form corresponding to GW radiation (see \citealt{AlcubierreLobo2017,BobrickMartire21}). If a warp drive spacetime that does have a GW signal far from the source were to be published, it would be interesting to include the signal in a search network.
\subsection{Near-Earth Trajectories}
\par
The \textsc{RAMACraft} considered in this study have been of astrophysical scale. We might, however, be interested in finding the detectable parameter space for near-Earth trajectories (NETs) from a distance within our solar system $R \leq 10^{16}$\,km, or even trajectories close to our atmosphere $R \leq 10^{6}$\,km. However, at these distances, the Newtonian portion of the strain, not the far-field quadrupole, will almost certainly dominate the signal. Therefore, to assess the detection possibilities of objects that come closer to Earth, one should examine the Newtonian contribution of these types of signals. We leave this for a future study.
\subsection{Future Horizons}
As new systems of interest are postulated and the GW catalog continues to grow, a very large number of filters will be required to search for all of these signals. Using conventional MF pipelines, this will quickly lead to intractable computational demands. However, this demand might be circumvented by training a neural network on a large filter set \citep{DeepSNR}. The neural network could then sift through large data sets without needing to implement each filter for every search.
\par 
On the other hand, if GW detectors become more sensitive such that burst detections of signals emitted from these advanced technologies become a possibility, researchers would have the opportunity to reverse-engineer those modes of transportation. This is because the shape of the GW signal is entirely dependent on the trajectory of the object. Thus, as a burst signal is detected, one can attempt to reason the qualities of the transportation mechanism present based on the shape of the GW signal.
\section{Conclusion}
\label{sec:Concl}
We have computed the ranges at which a linearly-moving \textsc{RAMAcraft} with constant acceleration would be detected by LIGO or LISA using either a burst or a MF search with a detection threshold of $(S / N)_{\rm det} = 8$. We find on the upper end of our results that Jupiter-mass \textsc{RAMAcraft} undergoing a change in velocity of $\Delta v \sim 0.3\,c$ are detectable from 10--100\,kpc, or up to within range of about every star in the Milky Way. On the lower end, we find that Moon masses undergoing $\Delta v \sim 0.3\,c$ are detectable from 1--10\,pc, or at and beyond the distance to the nearest stars in our Galaxy.
\par 
At the present moment, it may be safely stated that there have been no coincident bursts detected by LIGO above $(S / N)_{\rm det} =8$.
Because burst searches are indifferent to the signal shape, we can use our results for the LIGO burst search to constrain the occurrence of certain \textsc{RAMAcraft} trajectories (while LIGO was operating). In particular, we can rule out 10$^{-1}\,\rm M_{\odot}$ objects achieving $\Delta v \sim 0.3 c$ throughout the entire Galaxy, Jupiter-mass objects up to about 1\,kpc, and 10 Moon-mass objects up to about 1\,pc. 
Such strong constraints are enabled exclusively by the LIGO sensitivity. With MF searches and the installment of low-frequency detectors like LISA, DECIGO, and the BBO, we expect that these constraints can be made much more stringent.
\par 
To this point, our analysis shows that the
sensitivity to these objects can be increased through the development of low-frequency detectors thanks to the $f^{-1}$ scaling of the signal spectrum. In particular, we find that DECIGO and the Big Bang Observer (BBO) will increase the search volume for these objects by a factor of $10^{6}$, while both LISA and PTAs may also provide benefits for detecting \textsc{RAMAcraft} undergoing long acceleration periods. More generally, the shifting of any detector frequency band by a factor $\delta < 1$ will increase the detection range by $\delta^{-1}$ for burst searches and $\delta^{-1/2}$ for MF searches. The prospect of 
improving our sensitivity to these objects is therefore promising, and is furthermore coupled to low-frequency detector improvements that are of a particular interest to the fundamental physics community as well. 
\par 
Optimism aside, a candidate detection of a \textsc{RAMAcraft} signal should be treated with skepticism. Burst-like signals may also be caused by astrophysical sources that temporarily exceed the sensitivity curve. Therefore, MF detection will be more conclusive than burst searches. Even then, scrutiny should be applied to distinguish whether other natural sources, such as highly-eccentric BBH mergers, can produce similar waveforms.
\par 
GW detection is a sophisticated science, though it is still in its infancy. As the methodology is further developed, the sensitivity of detectors may become such that the detection of these objects is a regular occurrence. In this spirit, it would be interesting to complete a fully-fledged search for these objects. Our next papers will include gauging the sensitivity of objects closer to Earth, a simulated MF search to verify and improve the robustness of the templates, and an investigation of real detector data for these objects. We invite the scientific community to join us.
\section*{Acknowledgements}
We thank Jerry Tessendorf,  Lavinia Heisenberg, Shaun Fell, Alex Nitz from PyCBC, Thibault Damour, Brandon Melcher, Tito Dal Canton, Stephen Fairhurst, Edward Rietman, Haydn Vestal, Justin Feng, Christopher Helmerich, Jared Fuchs, and Toolchest for their extremely valuable comments and discussions at various stages of this work.

\appendix 
\section{GW Signal from a Newtonian Rocket}
\label{app:app_a}
Here we calculate the GW signal produced by a Newtonian rocket and compare the result to the simplifications made in Section \ref{ref:Sec2LinAcc}.
\par 
For a Newtonian rocket, the rocket contribution will still satisfy (\ref{eqn:gen_hzz}). The contribution from the exhaust, however, is slightly more complicated to calculate. To begin, we note that the energy density contributed by the exhaust of a discrete rocket in the detector frame of reference is simply a sum of the contributions of each individual ejected mass $dm$:
\begin{equation}
\label{eqn:discreterocket}
    T^{00}_{E}
    =
    \sum_{dm} dm \cdot
    \delta(x)
    \delta(y)
    \delta(z - z_{dm}(t))
\end{equation}
where $z_{dm}(t)$ is the trajectory of each mass $dm$. For a rocket that has been expelling mass for a time $t$, each mass ejected at time $t' \leq t$ with ejection velocity $u(t')$ (which may change between masses) has the trajectory
\begin{equation}
    z_{dm}(t)
    =
    \tilde{z}(t') - u(t')\cdot (t - t')
\end{equation}
where $\tilde{z}(t')$ is the position of the rocket at the time of ejection. Given a continuous rate of change $\dot{M}$, we can generalize the discrete case (\ref{eqn:discreterocket}) to a rocket that continuously expels mass:
\begin{equation}
    T^{00}_{E}
    =
    \int_{0}^{t}
    \left(-\dot{M} dt'\right)
    \delta(x)
    \delta(y)
    \delta\left\{
    z - \left(\tilde{z}(t') - u(t')(t-t')\right)
    \right\}
\end{equation}
Then to calculate the strain produced by the exhaust, we start with the quadrupole
\begin{equation}
    I_{ij}^{(E)}
    =
    \int x_{i}x_{j} d^{3}x
    \int_{0}^{t}
    \left(-\dot{M} dt'\right)
    \delta(x)
    \delta(y)
    \delta\left\{
    z - \left(\tilde{z}(t') - u(t')(t-t')\right)
    \right\} \\
\end{equation}
Then for finite $t$, we can swap the integration order by Fubini's Theorem and find that
\begin{equation}
    I_{zz}^{(E)}
    =
    \int_{0}^{t}
    \left(-\dot{M}dt'\right)
    \left[
    \tilde{z}(t') - u(t')(t - t')
    \right]^{2}
\end{equation}
and after applying the quadrupole formula, we have
\begin{equation}
    h_{zz}^{(E)}
    =
    \left(\frac{2G}{Rc^{4}}\right)
    \frac{d^{2}}{dt^{2}}
    \left\{
    \int_{0}^{t}
    \left(-\dot{M}dt'\right)
    \left[
    \tilde{z}(t') - u(t')(t - t')
    \right]^{2}
    \right\}
\end{equation}
To compute both derivatives, we use the following identity:
\begin{equation}
    \frac{d}{dt}
    \left\{
    \int_{a(t)}^{b(t)}
    g(t',t) dt'
    \right\}
    =
    \int_{a(t)}^{b(t)}
    \frac{\partial g}{\partial t}
    dt'
    +
    b'(t)
    g(b,t)
    -
    a'
    g(a,t)
\end{equation}
Then we find that the strain contribution from the exhaust is given by
\begin{equation}
\label{eqn:A8}
    \overline{h}^{(E)}_{zz}
    =
    \frac{2G}{Rc^{4}}
    \left\{
    -2
    \int_{0}^{t}
    \dot{M}(t')
    u^{2}(t')
    dt'
    +
    2\dot{M}(t)z(t)u(t)\right.
\end{equation}
\[
    \left. -
    \ddot{M}(t)z^{2}(t)
    -
    2\dot{M}
    z\dot{z}
    \right
    \}
\]
Looking again at the contribution of the rocket:
\begin{equation}
    \overline{h}^{(R)}_{zz}
    =
    \frac{2G}{Rc^{4}}
    \left\{
    2M\cdot
    \left(
    \dot{z}^{2}
    +
    z\ddot{z}
    \right)
    +
    4\dot{M}
    z
    \dot{z}
    +
    \ddot{M}z^{2}
    \right\}
\end{equation}
we see that the last two terms in (\ref{eqn:A8}) cancel manifestly and that the second term $2\dot{M}zu$ is canceled by Newton's Third Law:
\begin{equation}
    2z\cdot \frac{dp_{E}}{dt}
    =
    2\dot{M}uz
    =
    -2z\cdot 
    \left(
    \dot{M}\dot{z}
    +
    M\ddot{z}
    \right)
    =
    -2z\cdot 
    \frac{dp_{R}}{dt}
\end{equation}
Therefore, we see that all of the terms cancel except the first term from both contributions, which are additive since $\dot{M} < 0$. So, the two non-canceling contributions to the strain are due to the kinetic energy of the rocket and exhaust.
\par 
Finally, one might protest that the first term in (\ref{eqn:A8}) is not the change in kinetic energy of the exhaust and is missing a term from the product rule:
\begin{equation}
    \frac{d}{dt}\left\{M_{E}u^{2}\right\}
    =
    -\dot{M}u^{2}(t)
    +
    2M_{E}u\dot{u}
\end{equation}
where $\dot{M} = -\dot{M_{E}}$ and $M_{\rm E}$ is the total ejected mass. However, the cumulative kinetic energy of the exhaust is not given by $M_{E}(t) u^{2}(t)$ because $u$ may change between different ejections. By modeling the rocket as the continuum limit of a discrete rocket, we are \textit{defining} the rate of change of the exhaust kinetic energy as
\begin{equation}
    \frac{dK_{E}}{dt}
    =
    -\dot{M} u^{2}(t)
\end{equation}
Then
\begin{equation}
\begin{split}
    K_{E}(t)
    &=
    \int_{0}^{t}
    \left(
    -\dot{M}dt'
    \right)
    u^{2}(t')\\
    &
    =
    M_{E}(t)u^{2}(t)
    -
    2\int_{0}^{t}{M_{E}u\dot{u}dt'}
    \\
    &\neq 
    M_{E}(t)u^{2}(t)
\end{split}
\end{equation}
\section{T Scaling of the Strain Spectrum for a Given Acceleration and Change in Velocity}
\label{app:b}
\par
\begin{figure}
\centering
\includegraphics[width=1.0\linewidth]{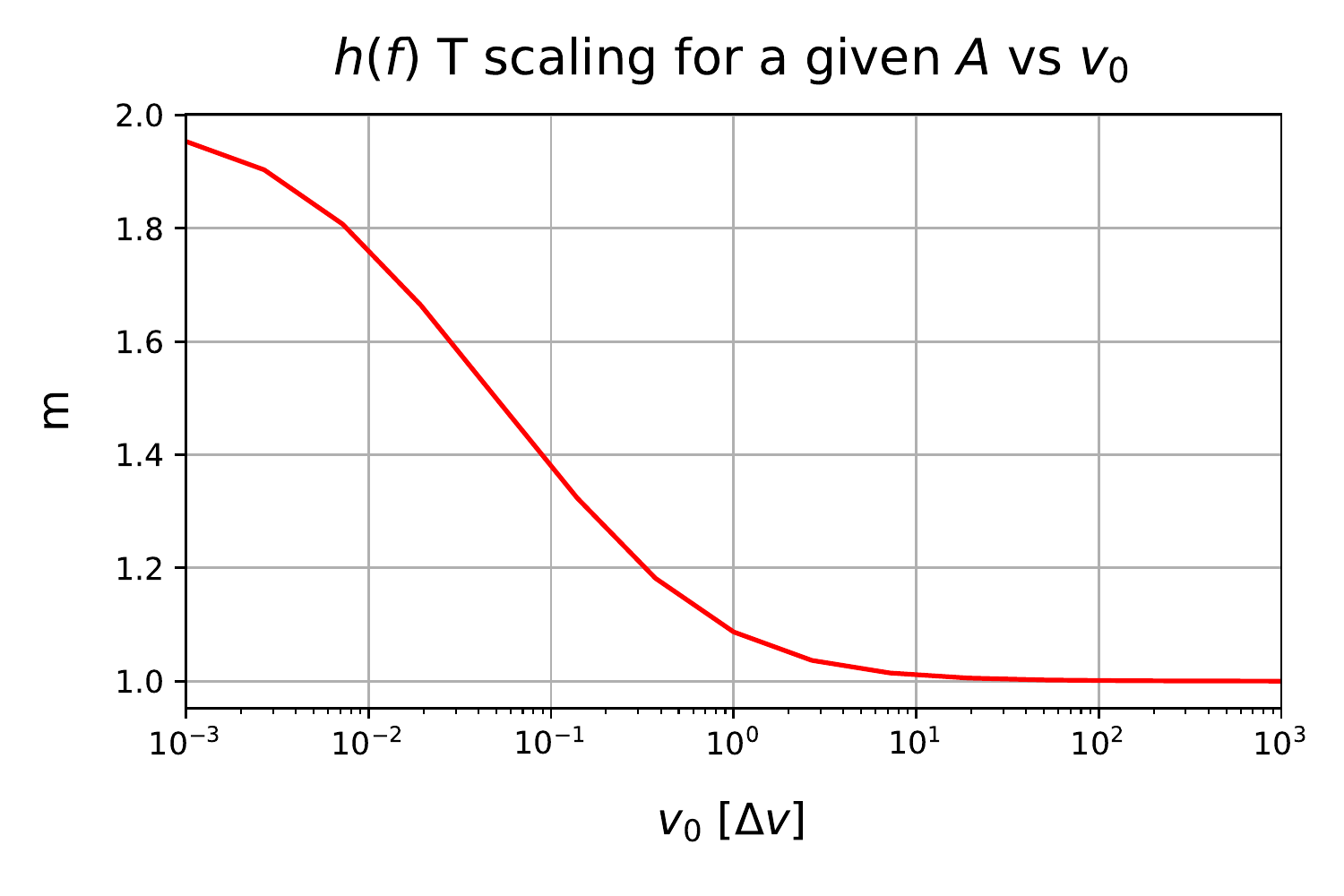}
\includegraphics[width=1.0\linewidth]{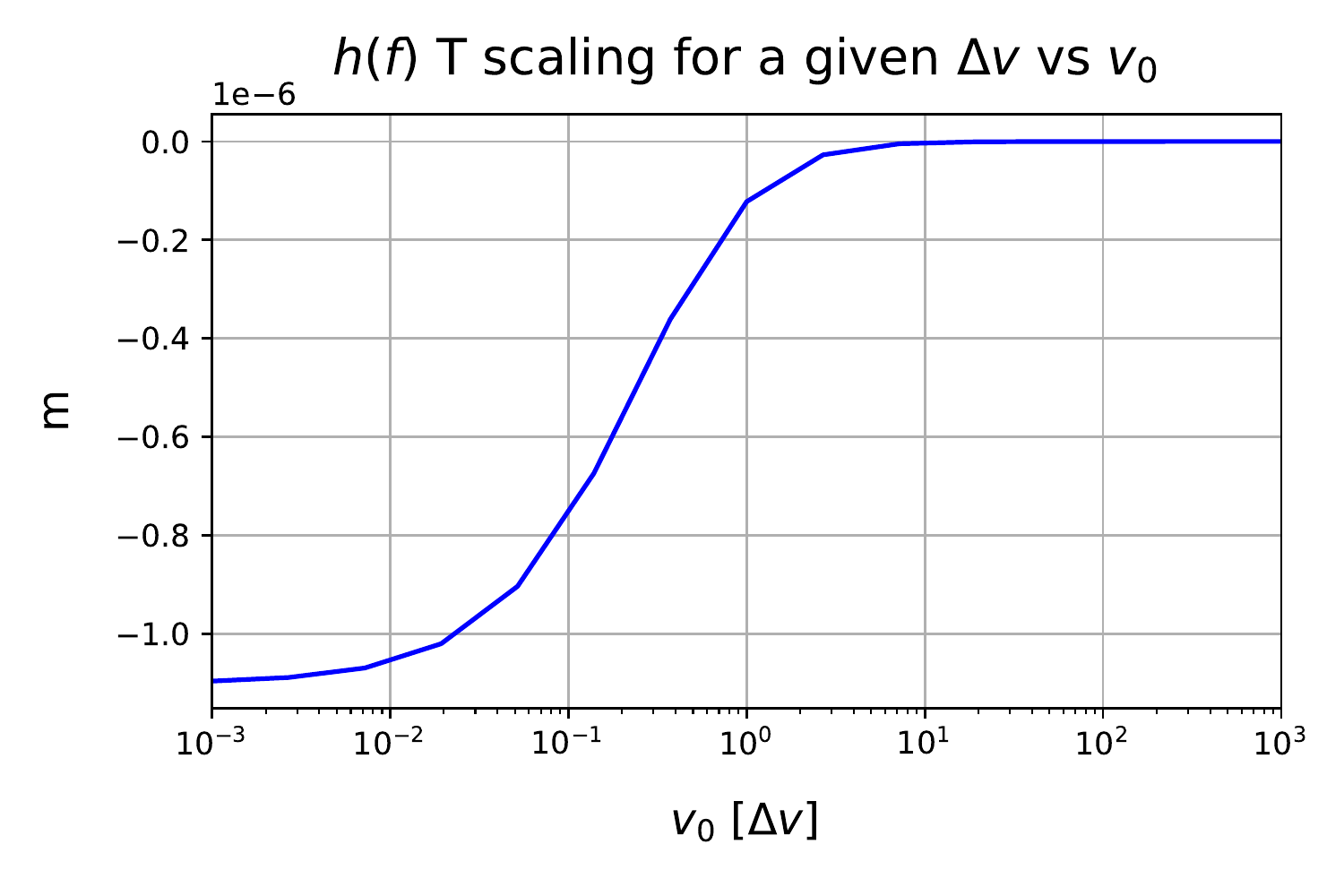}
\caption{Scaling behavior of the signal spectrum $h(f) \sim T^{m}$ for a given acceleration $A$ (Top) and change in velocity $\Delta v$ (Bottom). For a given $A$, we find that the scaling starts at $m = 2$ and decreases to $m = 1$, as expected. For a given $\Delta v$, we find that the scaling is relatively minuscule, as expected. The scaling does, however, increase for larger $v_{0}$.}
\label{fig:b1}
\end{figure}

 
Here we discuss the scaling of the signal spectrum when it is written for both a given acceleration $A$ (\ref{eqn:spectrum_A}) and a given change in velocity $\Delta v$ (\ref{eqn:spectrum_dv}). We parameterize the $T$ scaling by the following power law:
\begin{equation}
    h(f) \sim
    T^{m}
\end{equation}
and plot $m$ for different initial velocities $v_{0}$. We compute the curves by generating signal spectrums for a set of acceleration periods $T$ and compute the ratio of each spectrum with the spectrum computed using the smallest $T$ value. The scaling of this ratio yields the value of $m$, which we then compute for varying values of $v_{0}$. We compute each spectrum in the top curve in Figure \ref{fig:b1} with one acceleration $A$ and express the values $v_{0}$ in units of $AT_{f} = \Delta v$, where $T_{f}$ is the largest $T$ value used to compute the signal spectrums. We use this same set of $v_{0}$ for the bottom curve as well.
\par
For a given acceleration $A$, the spectrum should scale like $T^{2}$ for $\Delta v \gg v_{0}$ and $T$ for $v_{0} \gg \Delta v$. This is indeed what we see in Figure \ref{fig:b1}. The case for a given $\Delta v$ is less clear. If the $f^{-2}$ and $f^{-3}$ corrections to the $f^{-1}$ behavior are substantial, then we should find some $T$ dependence. In the bottom plot of Figure \ref{fig:b1}, since each $m$ is negative, we see that the spectrum (\ref{eqn:spectrum_dv}) decreases slightly with increasing $T$. However, the dependence is minuscule, which justifies our use of $h(f) \propto T^{0}$ for a given $\Delta v$ throughout the paper. 
\section{SNR vs Sample Length Scalings}
\label{app:d}
There are two pertinent results needed to explain the scaling behaviors (\ref{eqn:snrb_tau} -- \ref{eqn:MF_SNR_tau_bigT}). The first is the scaling of the noise correlation $\sigma_{\rm MF}$ with respect to the number of points used to compute the correlation sum. To see this effect, we plot the statistic $\sigma_{\rm MF}$ for a fixed mass, acceleration, acceleration period, and distance from the detector for varying $\tau \in T\cdot [1, 10^{3}]$. This choice of $\tau$ prevents the effect of a changing `perceived change in velocity' (\ref{SNRb_tau_bigT} -- \ref{eqn:MF_SNR_tau_bigT}) from interfering. 
\par
\begin{figure}
\centering
\includegraphics[width=1.0\linewidth]{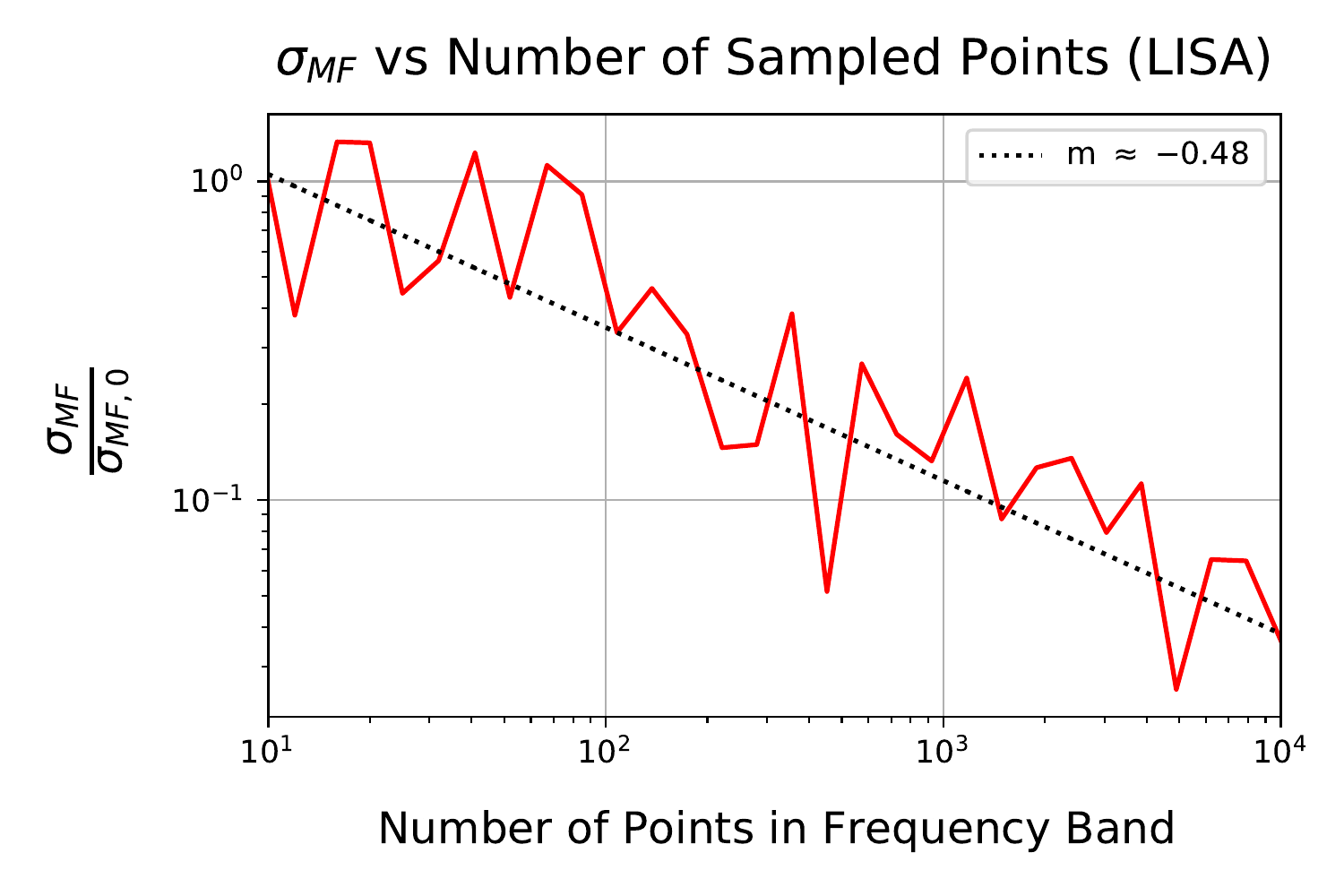}
\includegraphics[width=1.0\linewidth]{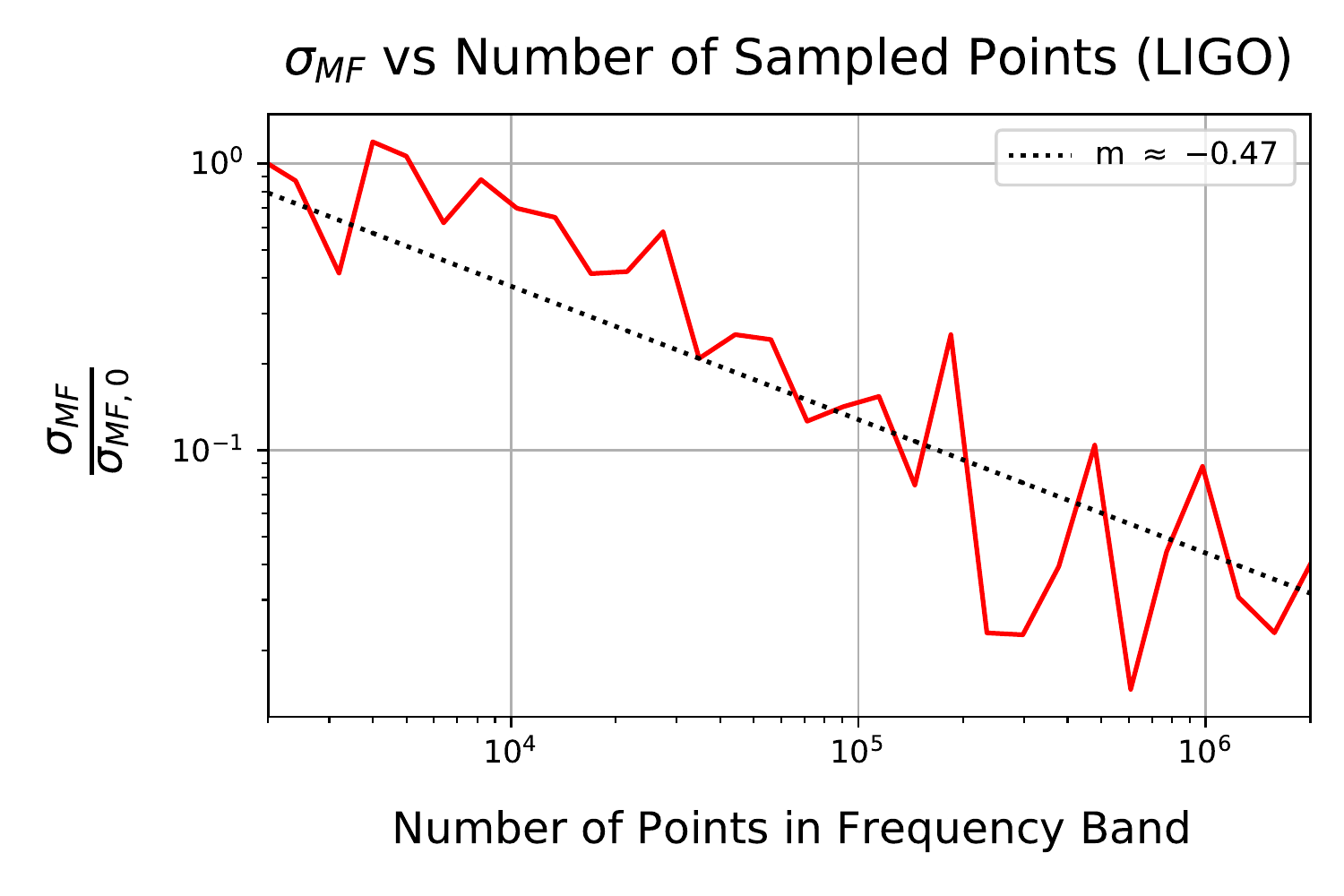}
\caption{Scaling behavior of the noise correlation $\sigma_{\rm MF}$ vs the number of points used to compute $\sigma_{\rm MF}$. We vary the number of points by varying $\tau$ for a fixed time spacing $dt$. We expect a contribution of $\tau^{-1/4}$, though we find an extra scaling factor of about $\tau^{-1/4}$. We conjecture that this is due to an increase in the number of points used to compute $\sigma_{\rm MF}$.}
\label{fig:d1}
\end{figure}

\begin{figure}
\centering
\includegraphics[width=1.0\linewidth]{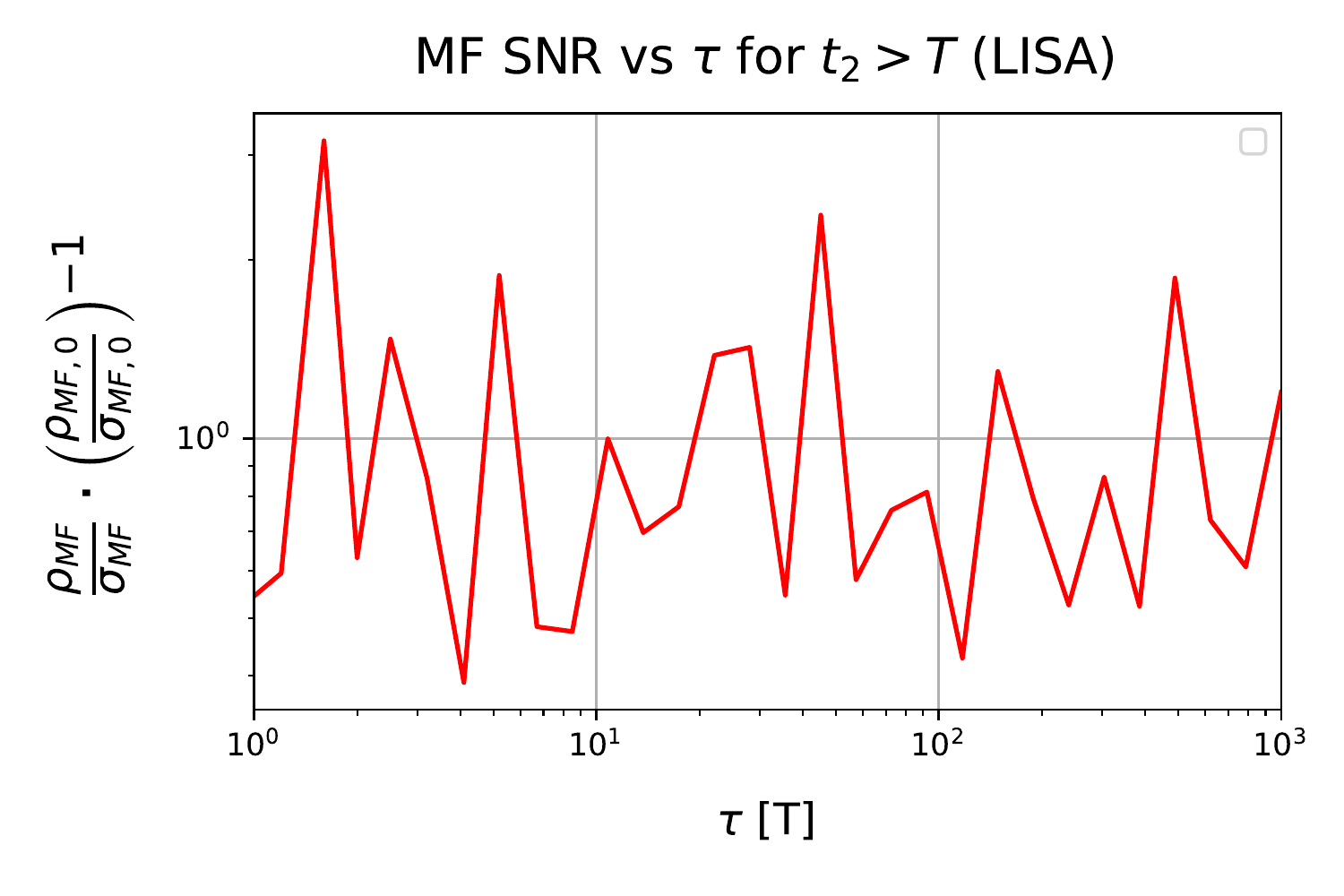}
\includegraphics[width=1.0\linewidth]{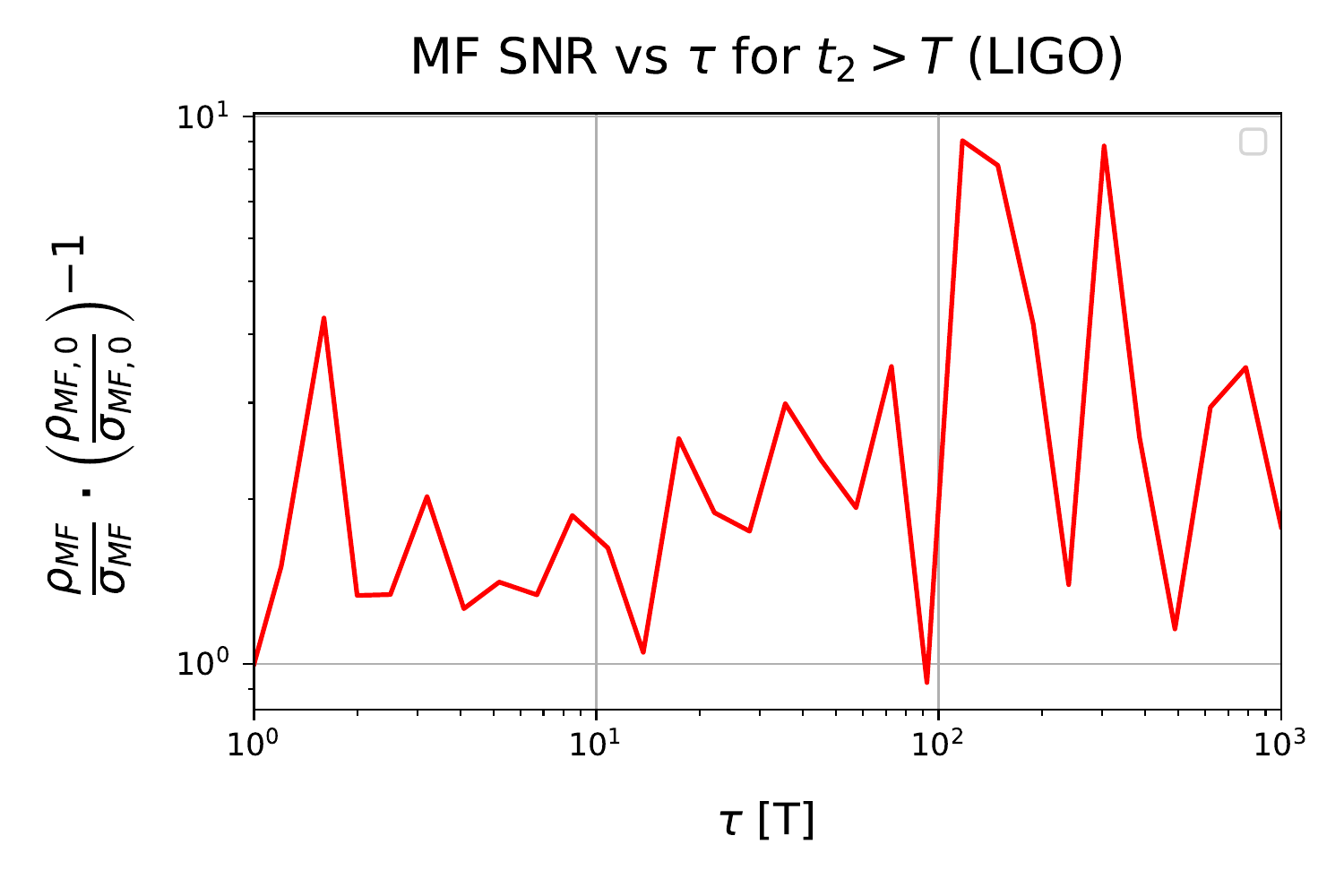}
\caption{Scaling behavior of the MF SNR $\rho_{\rm MF} / \sigma_{\rm MF}$  vs $\tau$ for the case where the acceleration period is contained entirely within the data sample. We find that the SNR is independent of $\tau$, as predicted in (\ref{eqn:MF_SNR_tau}). However, we also find that this prediction oscillates to within a factor of a few, which may be explained by the randomness involved in computing $\sigma_{\rm MF}$.}
\label{fig:d2}
\end{figure}

 
The results can be seen in Figure~\ref{fig:d1}, where we plot $\sigma_{\rm MF}$ in units of $\sigma_{\rm MF, 0}$, which is calculated with the initial sample length $\tau = T$. We expect a scaling of $\tau^{-1/4}$ for the correlation $\sigma_{\rm MF}$ based on (\ref{eqn:sigmf}), but there is an additional scaling of $\tau^{-1/4}$, combining to a total scaling of $\sigma_{\rm MF}\propto \tau^{-1/2}$. We conjecture that this is a result of the fact that larger $\tau$ values sample more points in a given frequency band by virtue of the decreased frequency spacing $\Delta f = \tau^{-1}$. So, $\sigma_{\rm MF}$ decreases with increasing $\tau$, and the SNR increases. This conjecture also agrees with the results in Sections \ref{sec:Results} and \ref{sec:Discussion}, as we state there as well. The result of this behavior is that the MF SNR is approximately independent of $\tau$, as can be seen in Figure~\ref{fig:d2} up to oscillations within a factor of a few. These oscillations are likely due to the randomness involved in computing the correlation (\ref{eqn:sigmf}). 
\par
\begin{figure}
\centering
       \includegraphics[width=1.0\linewidth]{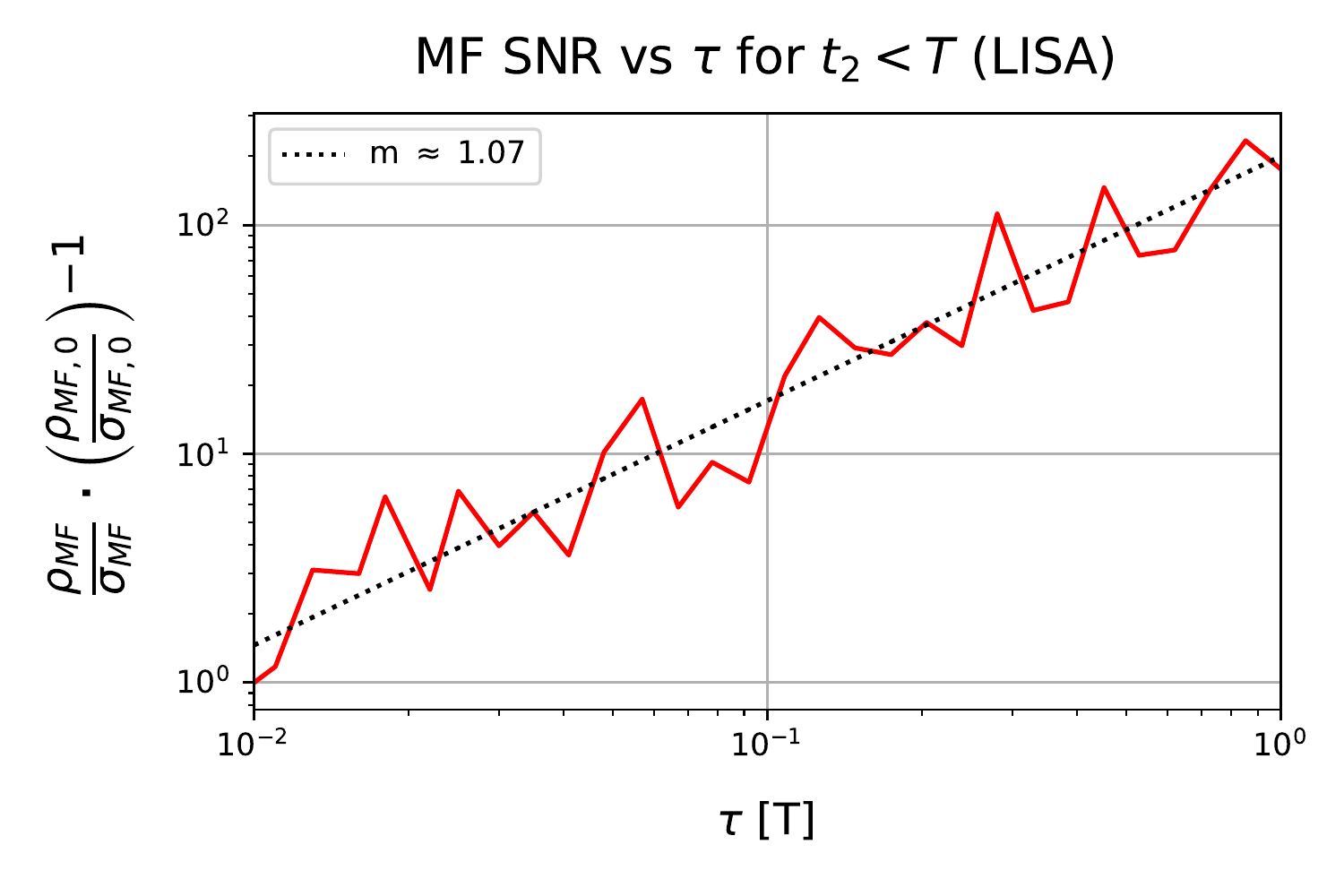}
       \includegraphics[width=1.0\linewidth]{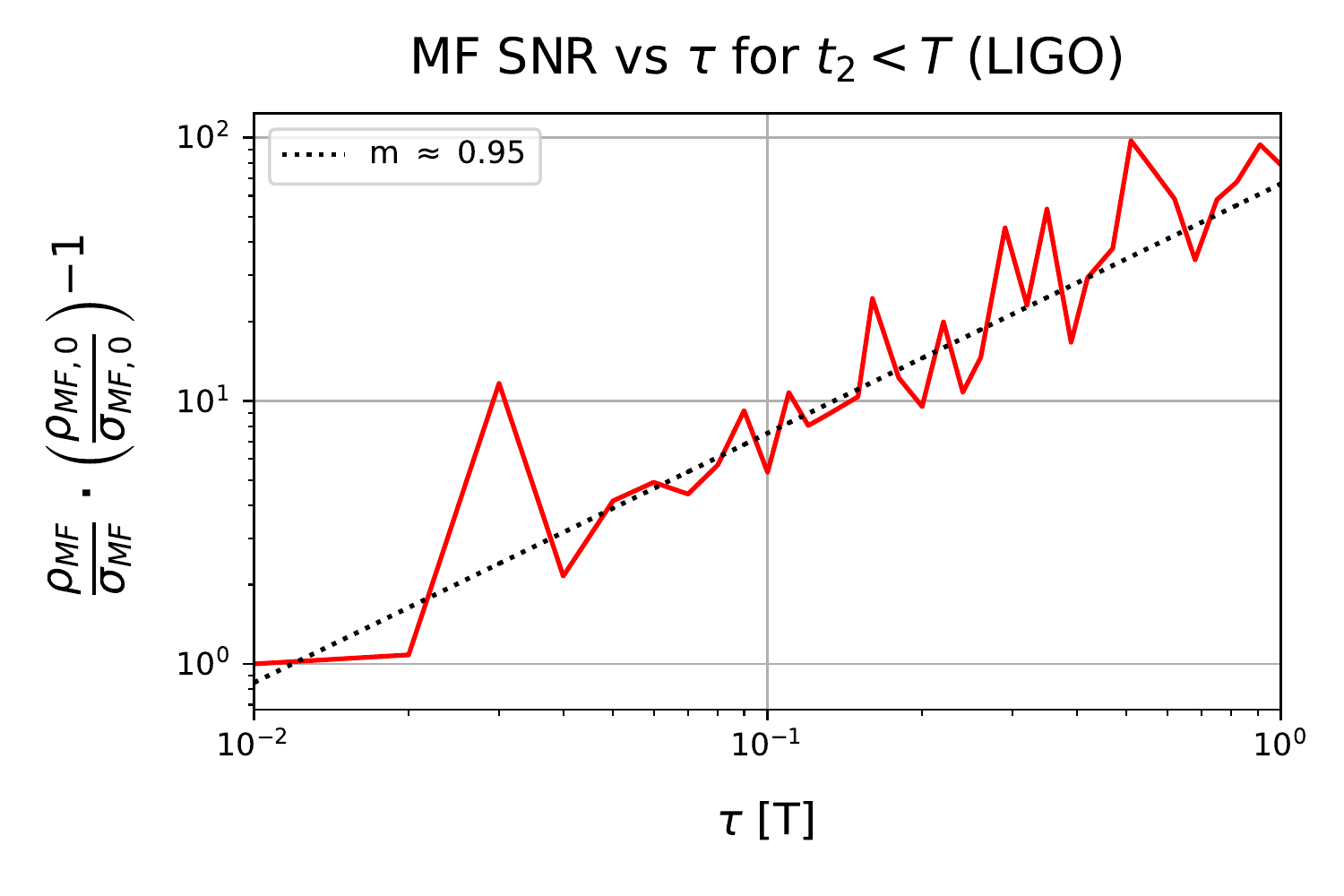}
    \caption{Scaling behavior of the MF SNR $\rho_{\rm MF} / \sigma_{\rm MF}$  vs $\tau$ for the case where the acceleration period extends past the data sample. We find that the SNR scales close to $\propto \tau^{1}$, as predicted in (\ref{eqn:MF_SNR_tau_bigT}).}
    \label{fig:d3}
\end{figure}

 
The second effect that explains the results (\ref{SNRb_tau_bigT} -- \ref{eqn:MF_SNR_tau_bigT}) is that for the case $t_{2} < T$, increasing $\tau$ will result in an increase in the `perceived change in velocity', which in turn increases the spectrum strength. In Figure~\ref{fig:d3}, we show that the resulting scaling for the MF SNR is $\rho_{\rm MF} / \sigma_{\rm MF} \propto \tau^{1}$, as predicted. 

%


\section*{Data availability}

The data underlying this article will be shared on reasonable request to the corresponding authors.

\clearpage
\bibliographystyle{mnras}
\bibliography{references}

\bsp	
\label{lastpage}
\end{document}